\documentclass[preprint,aps,superscriptaddress,floatfix,preprintnumbers,nofootinbib,altaffilletter]{revtex4-1}
\pdfoutput=1
\usepackage{epstopdf}
\usepackage{slashed}
\usepackage{graphicx} 
\usepackage{soul}
\usepackage{amsmath}
\usepackage{multirow}
\usepackage{tablefootnote}
\usepackage{comment}
\usepackage{xcolor}
\usepackage[utf8]{inputenc}
\usepackage{caption,subcaption}
\usepackage[normalem]{ulem}

\usepackage[T1]{fontenc}

\usepackage{graphicx}

\usepackage{amsmath}
\usepackage{amsfonts}
\usepackage{amssymb}
\usepackage{cleveref}
\usepackage{color}

\def\be{\begin{equation}}
\def\ee{\end{equation}}
\def\bea{\begin{eqnarray}}
\def\eea{\end{eqnarray}}

\def\<{\langle}
\def\>{\rangle}

\allowdisplaybreaks

\def\slashchar#1{\setbox0=\hbox{$#1$}           
   \dimen0=\wd0                                 
   \setbox1=\hbox{/} \dimen1=\wd1               
   \ifdim\dimen0>\dimen1                        
      \rlap{\hbox to \dimen0{\hfil/\hfil}}      
      #1                                        
   \else                                        
      \rlap{\hbox to \dimen1{\hfil$#1$\hfil}}   
      /                                         
   \fi}          

\def\mA{\mathcal{A}}

\def\mL{\mathcal{L}}

\def\mO{\mathcal{O}}

\def\bea{\begin{eqnarray}}
\def\eea{\end{eqnarray}}
\def\beq{\begin{equation}}
\def\eeq{\end{equation}}

\def\<{\langle}
\def\>{\rangle}

\def\agt{\gtrsim}

\begin{document}

\preprint{PITT-PACC-2324}

\title{Electroweak Scattering at the Muon Shot}

\author{Tao Han}
\affiliation{PITT PACC, University of Pittsburgh, Pittsburgh, PA, USA}

\author{Da Liu}
\affiliation{PITT PACC, University of Pittsburgh, Pittsburgh, PA, USA}
\affiliation{Center for Quantum Mathematics and Physics (QMAP),  University of California, Davis, CA 95616, USA}

\author{Ian Low}
\affiliation{High Energy Physics Division, Argonne National Laboratory, Argonne, IL 60439, USA}
\affiliation{Department of Physics and Astronomy, Northwestern University, Evanston, IL 60208, USA}

\author{Xing Wang}
\affiliation{Department of Physics, University of California at San Diego, La Jolla, CA 92093, USA}

\begin{abstract}
It has long been recognized that the scattering of electroweak particles at very high energies is dominated by vector boson fusion,  which probes the origin of  electroweak symmetry breaking and offers a unique window into the   ultraviolet regime of the SM. Previous studies assume SM-like couplings and rely on the effective $W$ approximation  (or electroweak parton distribution), whose validity is well-established within the SM but not yet studied in the presence of anomalous Higgs couplings. In this work, we critically examine the electroweak production of two Higgs bosons  in the presence of anomalous $VVh$ and $VVhh$ couplings. We compute the corresponding helicity amplitudes and compare the cross section results in the effective $W$ approximation with the full fixed-order calculation. In particular, we identify two distinct classes of anomalous Higgs couplings, whose effects are not captured by vector boson fusion and effective $W$ approximation. Such very high energy electroweak scatterings can be probed at the Muon Shot, a multi-TeV muon collider upon which we base our study, although similar considerations apply to {other high energy} colliders. 
\end{abstract}

\maketitle

\noindent
\section{Introduction}

With the discovery of the 125 GeV Higgs boson in 2012 at the Large Hadron Collider (LHC), the Standard Model (SM) of particle physics is a UV-consistent theory. Although the Higgs boson is often hailed as the origin of mass for (almost) all fundamental particles, a key feature of the SM Higgs lies in the fact that it unitarizes the electroweak vector boson scattering. In particular, if couplings of the 125 GeV Higgs with the electroweak vector bosons deviate from the SM predictions even just a tiny bit, the amplitude for vector boson scattering would grow with energy and eventually violates perturbative unitarity.

\begin{figure}[tb]
\centering
\includegraphics[width=0.25\textwidth]{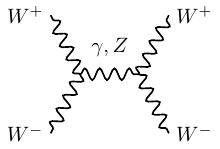}
\includegraphics[width=0.25\textwidth]{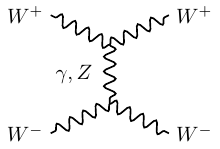}
\includegraphics[width=0.25\textwidth]{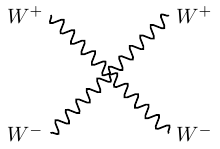}\\
\includegraphics[width=0.25\textwidth]{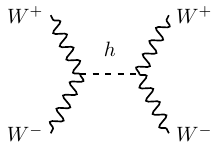}
\includegraphics[width=0.25\textwidth]{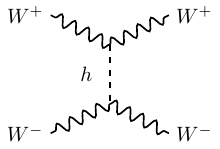}
\caption{\em SM Feynman diagrams contribute to the scattering $W^+W^-\to W^+W^-$. }
\label{fig:vvscattering}
\end{figure}

Consider the 2-to-2 scattering $W^+W^-\to W^+W^-$ in the SM as shown in Fig.~\ref{fig:vvscattering}. Besides the triple and quartic  gauge interactions,  it includes the Higgs boson $(h)$ as an intermediate particle. 
In the absence of the Higgs contribution, the leading behavior of the amplitude from the top row has the parametric dependence \cite{Peskin:1995ev}
\be
\label{eq:wwscat}
{\cal M}(W^+W^-\to W^+W^-) \sim \frac{s}{v^2} \ ,
\ee
where $s=(p_1+p_2)^2$ is the centre-of-mass energy squared and $v$ is the vacuum expectation value of the Higgs field. The energy growth eventually leads to unitarity violation. This  behavior can be seen explicitly from  the polarization vectors for the transversely (T) and longitudinally (L) polarized $W$ boson along the $\hat{z}$-axis as 
\be
\epsilon^\mu_T = \left(0,\pm \frac1{\sqrt{2}}, -\frac{i}{\sqrt{2}}, 0\right) \ , \qquad \epsilon^\mu_L = \frac1{m_W}\left(|\vec{k}|, 0,0,E_W\right) \ ,
\ee
where $\vec{k}$ is the three-momentum  and $E_W=\sqrt{|\vec{k}|^2+m_W^2}$ is the energy of the $W$ boson. In the high energy limit, $E_W\gg m_W$, $\epsilon_L^\mu \approx k^\mu/m_W + {\cal O}(m_W/E_W)$, thus 
\be
\epsilon_T^{W^+}\cdot \epsilon_T^{W^-} \sim 1,\qquad 
\epsilon_L^{W^+}\cdot \epsilon_L^{W^-} \approx \frac{k_{W^+}\cdot k_{W^-}}{m_W^2} \sim \frac{s}{2m_W^2} \ .
\ee
Although the leading behavior of each individual diagram is $(\epsilon_L^{W^+}\cdot \epsilon_L^{W^-})^2 \sim s^2/m_W^4$, the gauge invariance guarantees that terms proportional to $s^2$ cancel and only the linear term in $s$ remains  when all diagrams in the top row are included. In the absence of the Higgs, the linear growth in $s$ in Eq.~(\ref{eq:wwscat}) is completely analogous to the pion-pion ($\pi\pi$)  scattering near the threshold in low-energy QCD. There the energy growth is due to the fact that Nambu-Goldstone bosons are derivatively coupled~\cite{Cornwall:1974km,Lee:1977eg, Chanowitz:1985hj,Vayonakis:1976vz,Low:2014nga,Low:2018acv}.

The analogy between $WW$ scattering and pion-pion scattering actually highlights a great mystery in the SM: in QCD the $\pi\pi$ scattering is partially unitarized by a {\em series} of resonances until the QCD confinement scale $\sim1$ GeV, including the spin-1 $\rho$ meson, while the $WW$ scattering is unitarized by a single scalar particle -- the Higgs boson. Why is there such a distinction? More importantly, using a single particle to unitarize the $WW$ scattering means couplings of the Higgs with the vector bosons, the $VVh$  coupling, must have the exact form and strength as predicted by the SM. If the $VVh$ coupling deviates from the SM even just by a small amount,  the cancellation  would be incomplete and an energy growing term in Eq.~(\ref{eq:wwscat}) reappears.  Pion-pion scattering in low-energy QCD, to the contrary, is unitarized sequentially by a tower of  resonances, each of which pushes the scale of unitarity violation further  to a higher energy, eventually reaching above the scale of chiral symmetry breaking. When chiral symmetry is restored, pions cease to exist. In the Standard Model, the 125 GeV Higgs alone would unitarize VV scattering up to an arbitrarily high scale. This was one of the clearest indicators on the presence of a Higgs-like particle before the discovery of the 125 GeV Higgs. Experimentally, at the LHC, we have measured the HVV coupling to be consistent with the SM expectation up to ${\cal O}(10\%)$ uncertainty, which suggests the 125 GeV Higgs is responsible for unitarizing VV scattering up to 10 TeV. Nevertheless, for such a critical prediction of the SM, it is important to continue to investigate whether the 125 GeV Higgs could unitarize VV scattering up to an even higher energy scale.

\begin{figure}[tb]
\centering
\begin{subfigure}{0.2\textwidth}
\includegraphics[width=\textwidth]{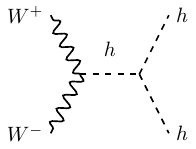}
\caption{}
\label{fig:feyn_s}
\end{subfigure}
\hspace{0.2cm}
\begin{subfigure}{0.2\textwidth}
\includegraphics[width=\textwidth]{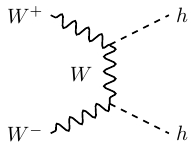}
\caption{}
\label{fig:feyn_t}
\end{subfigure}
\hspace{0.2cm}
\begin{subfigure}{0.2\textwidth}
\includegraphics[width=\textwidth]{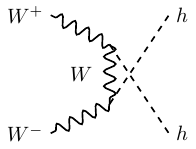}
\caption{}
\label{fig:feyn_u}
\end{subfigure}
\hspace{0.2cm}
\begin{subfigure}{0.2\textwidth}
\includegraphics[width=\textwidth]{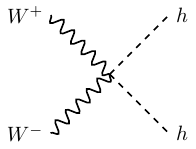}
\caption{}
\label{fig:feyn_4}
\end{subfigure}
\caption{\em Feynman diagrams contribute to the scattering $W^+W^-\rightarrow h h$. From left to right they are the $s$-, $t$-, $u$-channel and the 4-point interaction channel.}
\label{fig:feyn}
\end{figure}

These arguments are the reason why vector boson scattering, or vector boson fusion (VBF), is among the top priorities in current and future experimental programs at a high energy collider. In this work we will focus on the production of two Higgs bosons at a very high energies, which in the SM is dominated by the sub-process $VV\to hh$. This process is of particular importance for several reasons, in addition to what has already been articulated. In the SM $VV\to hh$ involves diagrams shown in Fig.~\ref{fig:feyn}, which contains both the four-point $VVhh$ coupling and the trilinear $hhh$ coupling, neither of which have been measured experimentally. The $VVhh$ coupling arises from the Higgs kinetic term,
\be
\label{eq:vvhhsm}
D^\mu H^\dagger D_\mu H \supset \frac{h^2}{v^2}\left(m_W^2 W_\mu^- W^{\mu\,+} + \frac12 m_Z^2 Z_\mu Z^\mu\right) \ ,
\ee
where $H$ is the Higgs doublet. While the trilinear $hhh$ coupling is part of the Higgs potential that triggers the electroweak symmetry breaking,
\be
V(H) \supset - \frac{m_h^2}{2v} h^3\ ,
\ee
where $v=246$ GeV is the vacuum expectation value (VEV) of the Higgs. Therefore $VV\to hh$ allows us to probe three important aspects of the SM: 1) unitarity in VBF, 2) gauge invariance in the Higgs sector, and 3) the shape of the Higgs potential, all of which have yet to be verified experimentally.

Recently there has been significant interest  in a multi-TeV muon 
collider\footnote{The program towards a multi-TeV muon collider is coined the ``Muon Shot'' in Ref.~\cite{P5}.} \cite{Delahaye:2019omf,MuonCollider:2022xlm,Accettura:2023ked}, which would offer a direct access to the high energy behavior of the SM and the potential discovery for BSM new physics.  
At an energy scale much larger than the electroweak scale, all relevant degrees of freedom become light and the probability for an energetic electroweak particle to emit an electroweak gauge boson is enhanced by the collinear singularity, which is regulated by the non-zero masses and manifest itself through the collinear logarithms. At the same time it is a good approximation to treat the electroweak gauge bosons as on-shell particles, which is known as the effective $W$ approximation (EWA) \cite{Dawson:1984gx,Kane:1984bb,Ruiz:2021tdt}. This formalism has been further developed to a partonic picture of the electroweak (EW) interactions in high energy collisions  \cite{Ciafaloni:2005fm,Chen:2016wkt,Bauer:2017isx,Han:2020uid}. 
A very high energy lepton collider, such as a multi-TeV muon collider, thus serves as a vector-boson collider 
\cite{Buttazzo:2018qqp,Costantini:2020stv,Han:2020uid,AlAli:2021let}. 
The cross section for $VV\to hh$ in leptonic collisions was first calculated in Ref.~\cite{Barger:1988kb}, and   subsequently studied in  \cite{Contino:2010mh,Kilian:2018bhs,Han:2020pif}. However, previous studies always assume SM tensor structures in Eq.~(\ref{eq:vvhhsm}) and only modify the coupling strength.  We will not make the same assumption and instead will consider the possibility of anomalous Higgs coupling in a general framework of nonlinear effective field theory (EFT) \cite{Low:2014oga,Liu:2018vel,Liu:2018qtb,Liu:2019rce}. Allowing for anomalous couplings offers a unique opportunity to understand the dominance of VBF in electroweak scattering, as well as the associated EWA, in a broader context when the 125 GeV Higgs does not completely unitarize the vector boson scattering.  Indeed, we will see that, after including anomalous couplings,  there are effects not captured by the VBF and EWA, and a full fixed order calculation is warranted. 

The rest of the paper is organized as follows. In Section~\ref{sec:eft}, we set up our effective field theory notation, and discuss the linear and nonlinear realization of the Higgs boson. In Section~\ref{sec:ww}, we give the scattering  amplitudes of the $W$ boson pair into a pair of Higgs boson, and discuss the threshold and the high energy limit behaviors. In Section~\ref{sec:ewabsm}, we analyze the double Higgs production at a multi-TeV muon collider and discuss the validity of the EWA in the context of EFT. In Section~\ref{sec:kinematics}, we discuss the kinematic features of the double Higgs production at muon colliders. We conclude in Section~\ref{sec:conclusion}. The Feynman rules for the anomalous Higgs-$W$ couplings and the general helicity amplitudes are given in two appendices. 

\section{EFT: Linear versus Nonlinear Realization}
\label{sec:eft}

An effective field theory (EFT) by construction consists of an infinite number of operators with increasing powers of derivatives and fields. Using scalar fields as an illustration and denoting by $\Phi=\{\phi_1,\phi_2,\cdots\}$ a generic set of scalars, an EFT is a double expansion in 
\be
\label{eq:powerc}
\frac{\partial_\mu}{\Lambda}  \qquad \text{and} \qquad \frac{\Phi}{f} \ ,
\ee 
where  $\Lambda$ and $f$ are two mass scales characterizing the momentum expansion and the field expansion. Then the effective Lagrangian has the following general form
\be
\label{eq:effLgen}
{\cal L}_{eff} = \Lambda^2 f^2\ \tilde{\cal L}(\partial/\Lambda, \Phi/f) \ ,
\ee
where dimensionless coefficients in $\tilde{\cal L}$ are assumed to be order unity and the overall factor is fixed by requiring a canonically normalized kinetic term, which contains two derivatives and two scalars. Then Naive Dimensional Analysis (NDA) \cite{Manohar:1983md,Luty:1997fk,Cohen:1997rt} indicates that loop effects below the scale $\Lambda$ are suppressed by a factor of
\be
\label{eq:loopfactor}
L=\frac{1}{16\pi^2} \left(\frac{\Lambda}{f}\right)^2 \ ,
\ee
when comparing to the tree-level effect. This suggests defining $\mathsf{g}=\Lambda/f$ as some sort of ``coupling constant'' in the EFT.  The effective Lagrangian $\tilde{\cal L}$ in Eq.~(\ref{eq:effLgen})  now has the following structure,
\be
\tilde{\cal L} = \tilde{\cal L}^{(0)}(\partial/\Lambda, \Phi/f)+ \frac{\mathsf{g}^2}{16\pi^2}\, \tilde{\cal L}^{(1)}(\partial/\Lambda, \Phi/f) +\frac{\mathsf{g}^4}{(16\pi^2)^2}\, \tilde{\cal L}^{(2)}(\partial/\Lambda, \Phi/f)  +\cdots  \ .
\ee
Then, in a weakly-coupled theory characterized by
\be 
\mathsf{g} \ \lesssim\  {\cal O}(1)  \qquad \text{and} \qquad \Lambda\  \sim\ f\ ,
\ee
one does not distinguish between $\Lambda$ and $f$ in the EFT. This is the assumption underlying a linear EFT such as the Standard Model Effective Field Theory (SMEFT) \cite{Buchmuller:1985jz,Leung:1984ni,Hagiwara:1993ck,Grzadkowski:2010es}.  On the other hand, if the effective theory becomes strongly-coupled at a certain scale and
\be 
\mathsf{g} \ \lesssim\  4\pi  \ ,
\ee
then there could be a distinction between $\Lambda$ and $f$. However, the separation could at most be
\be
 \Lambda\  \sim\ 4\pi f \ ,
 \ee
where $\mathsf{g} \sim 4\pi$.
In this case the EFT loses predictive power at the energy scale $\sim 4\pi f$ and needs to be UV-completed. This is usually referred to as a nonlinear realization of the EFT.\footnote{See Ref.~\cite{Falkowski:2019tft} for a recent classification of the effective field theories (linear versus non-linear) based on the analyticity of the Lagrangian with respect to the Higgs doublet.} The most well-known example  arises from the chiral Lagrangian in low-energy QCD \cite{Weinberg:1996kr}.  Another example is the composite Higgs model \cite{Kaplan:1983fs,Kaplan:1983sm}, where the Higgs boson is a pseudo-Nambu-Goldstone boson (pNGB), and its modern incarnations \cite{ArkaniHamed:2001nc,ArkaniHamed:2002qx,ArkaniHamed:2002qy,Contino:2003ve,Agashe:2004rs,Panico:2015jxa,Barnard:2013zea, Ferretti:2014qta,Erdmenger:2020flu}.\footnote{Strictly speaking, a pNGB Higgs does not necessarily imply a ``composite'' Higgs in the sense that the proton is composite, because one could UV-complete a pNGB Higgs into a weakly-coupled description a l\'{a} the linear sigma model. (See, for example, Ref.~\cite{Csaki:2008se}.)
}

A weakly-coupled EFT makes qualitatively different low-energy predictions from a strongly-coupled EFT. For example, given the discovery of the 125 GeV Higgs boson, it is commonly accepted that  the electroweak $SU(2)_L\times U(1)_Y$ symmetry is linearly realized. Then in a weakly-coupled EFT such as the SMEFT, where one does not distinguish between the derivative expansion $\partial_\mu/\Lambda$ and $\Phi/f$, the leading corrections to $VVh$ and $VVhh$ vertices come from dim-6 operators and they are correlated,
\be
\label{eq:smefthvv}
\frac{1}{\Lambda^2} (H^\dagger H)*\mathsf{O}_{VV} \ \to \  \frac{v^2}{2\Lambda^2} \left(\frac{h^2}{v^2}+ \frac{2h}{v}+1\right)* \mathsf{O}_{VV} \ ,
\ee
where the vacuum expectation value $\langle H \rangle =(0, v)^T/\sqrt{2}$ and $\mathsf{O}_{VV}$ represents a dim-4 operator containing two electroweak gauge bosons $V=\{W, Z, \gamma\}$. 
Therefore, once we measure $VVh$  couplings, the quartic $VVhh$  couplings are also known in SMEFT, until one further introduces dim-8 operators. But of course if the power counting of the EFT is well-defined, the effect of dim-8 operators must be smaller than those from the dim-6 operators.

This strong correlation between $VVh$ and $VVhh$ vertices is not present in a strongly-coupled EFT like in the composite Higgs models. The reason is because one could be making a measurement at an energy scale $E$ such that
\be
\frac{v}{f}\ \agt\ \frac{E}{\Lambda} ,
\ee
 in which case we would need to include corrections that are to all orders in $1/f$ expansion, and organize the power-counting by the $\partial/\Lambda$ expansion. For example, at the leading two-derivative order we could have, in the unitary gauge,\footnote{Recall the each gauge boson is counted as one derivative, because of covariant derivative $D_\mu = \partial_\mu- i e A_\mu$.}
 \be
 \mathsf{F}(h/f) * \left(m_W^2 W^+_\mu W^{-\mu}+\frac{1}{2}m_Z^2 Z_\mu Z^{\mu}\right)\ ,
 \ee 
 where $\mathsf{F}$ is an analytic function resumming all the $1/f$ effect  and $m_{W/Z}$ is the mass of the $W/Z$ boson. Here we have normalized in a way such that $\mathsf{F}(0) = 1$. We have also assumed the $SU(2)_C$ custodial invariance in the Higgs sector. After electroweak symmetry breaking and $h\to h+v$, the corrections to $VVh$ and $VVhh$  vertices are given by
 \be
 \label{eq:nonlinHVV}
  \left[\mathsf{F}^\prime(0)\ \frac{h}{f}+ \mathsf{F}^{\prime\prime}(0) \frac{h^2}{f^2}  \right]* \left(m_W^2 W^+_\mu W^{-\mu}+\frac{1}{2}m_Z^2 Z_\mu Z^{\mu}\right)\ .
 \ee
One sees explicitly that, unlike in SMEFT,  corrections to $VVh$ and $VVhh$  vertices are now independent, governed by the underlying theory encoded in $\mathsf{F}$. 

In this work we would like to consider the more general possibility that $VVh$  and $VVhh$ vertices are not correlated, and thus focus on the pNGB Higgs models. There are dozens of pNGB Higgs models in the literature \cite{Bellazzini:2014yua} and they differ in the choices of an extended (approximate) global symmetry $\mathsf{G}$ in the UV which is spontaneously broken down to a subgroup $\mathsf{H}\supset SU(2)_L\times U(1)_Y$ in the IR. For a phenomenologically successful model, it is important to impose the custodial invariance in the Higgs sector and $\mathsf{H}\supset SO(4)$ \cite{Bellazzini:2014yua} and the four real components of the Higgs doublet transform as the fundamental representation under $SO(4)$. We adopt this assumption here.

For a pNGB Higgs, the  function $\mathsf{F}(h/f)$ turns out to have some very interesting properties that are not realized until recently. In explicit models, $\mathsf{F}(h/f)$ looks seemingly different for different choices   of  symmetry breaking pattern $\mathsf{G}/\mathsf{H}$. However, it was discovered recently that $\mathsf{F}(h/f)$ actually only depends  on the IR quantum number of the Higgs boson \cite{Low:2014nga,Low:2014oga}, up to the normalization of $f$. So in models where the Higgs doublet belongs to the fundamental representation of an $SO(4)$ subgroup of $\mathsf{H}$, $\mathsf{F}(h/f)$ is in fact {\em universal} among different $\mathsf{G}/\mathsf{H}$, after a suitable re-scaling of $f$, and has the form after electroweak symmetry breaking
\bea
\mathsf{F}(h/f)&=&
\frac{f^2}{v^2} \left(\frac{v}f \ \cos\frac{h}{f}+\sqrt{1-v^2/f^2}\ \sin\frac{h}{f}\right)^2 \nonumber \\
 &=& \frac{f^2}{v^2} \sin^2(\theta+h/f) \ ,
 \eea
 where 
 $\sin\theta \equiv  v/f = \sqrt{\xi}$. This will give rise to the ratio as follows:
 \beq 
 \frac{\mathsf{F}^{\prime \prime} (0)}{\mathsf{F}^{\prime}(0)} = \frac{2 \sqrt{\xi} \sqrt{1 - \xi}}{1 - 2 \xi}.
\eeq 
The only free parameter is then $f$, the Goldstone boson decay constant, which needs to be determined  experimentally as a theory input. It is worth emphasizing that, although at the two-derivative order, the function $\mathsf{F}(h/f)$ is determined from the infrared, higher-derivative contributions, say ${\cal O}(p^4)$ corrections, it encodes the unknown UV physics through the uncalculable Wilson coefficients, as is common in EFTs.

In composite Higgs models, corrections to $VVh$ and $VVhh$ interactions that are next-to-leading order in the derivative expansion were enumerated in Refs.~\cite{Liu:2018vel,Liu:2018qtb}, following which we write down the following nonlinear effective Lagrangian involving $WWh$ and $WWhh$  vertices up to ${\cal O}(p^4)$, in the unitary gauge
\begin{equation}
\label{eq:nonL}
\aligned
\mathcal{L}_{\rm EFT} =&~ \mathcal{L}_{\rm SM} + \left(2C_0^h \frac{h}{v} +  C_0^{2h}\frac{h^2}{v^2}\right)\left(m_W^2 W^+_\mu W^{-\mu}+\frac{1}{2}m_Z^2 Z_\mu Z^{\mu}\right)\\
&~ + C^h_5\left( \frac{h}{v}W^+_\mu\mathcal{D}^{\mu\nu}W^-_\nu + {\rm h.c.}\right) + C^h_6\frac{h}{v}W^+_{\mu\nu}W^{-\mu\nu} \\
&~ + C^{2h}_5\left(\frac{h^2}{v^2}W^+_\mu\mathcal{D}^{\mu\nu}W^-_\nu +{\rm h.c.}\right) + C^{2h}_6\frac{h^2}{v^2}W^+_{\mu\nu}W^{-\mu\nu} \\
&~ + C^{2h}_9\frac{(\partial_\nu h)^2}{v^2}W^+_\mu W^{-\mu} + C^{2h}_{10}\frac{\partial^\mu h\partial^\nu h}{v^2}W^+_\mu W^-_\nu,
\endaligned
\end{equation}
where $\mathcal{D}^{\mu\nu} = \partial^\mu\partial^\nu-\eta^{\mu\nu}\partial^2$. 
Here and henceforth, we choose to normalize the higher dimensional operators to the Higgs VEV $v$.
For reader's convenience, we provide the Feynman rules for the interacting vertices in Appendix~\ref{appendix:B}. 
The Wilson coefficients $C_i^h$ and $C_i^{2h}$ are assumed to be independent of each other, and both enter the electroweak double Higgs production.\footnote{We focus on $WW\to hh$ in this work. It would be desirable  to include $ZZ\to hh$ in future studies, due to the inability to distinguish the two channels in a very high energy collider \cite{Han:2020pif}.}
Eq.~(\ref{eq:nonL}) makes it convenient to compare with experimental observables \cite{Liu:2019rce}. However, the power counting rule in Eq.~(\ref{eq:powerc}) is not manifest in the unitary gauge and the natural size of $C_i$'s is not order unity. This motivates the following rescaling,
\be
\label{Eq:powerCC}
\aligned
C_i^{h/2h} &\to \widetilde{C}_i^{h/2h}*\frac{v^2}{f^2} \ , \ \ i=0, 5, 6 \ ; \quad \ \ 
C_j^{h/2h} &\to \widetilde{C}_j^{h/2h}*\frac{v^4}{f^4} \ , \quad j=9, 10 \ ,
\endaligned
\ee
which makes it clear that operators multiplied  by $C_i^{h/2h}$ are matched to dim-6 operators in SMEFT at the leading order, while those multiplied by $C_j^{h/2h}$ are matched to dim-8 operators. After the re-scaling, $\tilde{C}^{h/2h}$'s are expected to be order unity. Note that the coupling coefficients can be written as the Wilson coefficients of the SMEFT operators as follows~\cite{Liu:2018vel,Liu:2018qtb}:
\beq
\begin{split}
C_0^{2h} &= 4 C_0^h = -\frac{v^2}{2f^2} c_H\\
C_5^{2h} &= \frac 12 C_5^h = 2 \frac{m_W^2}{m_\rho^2} (c_W + c_{HW}), \qquad C_6^{2h} = \frac 12 C_6^h = -2 \frac{m_W^2}{m_\rho^2}  c_{HW} \\
C_9^{2h} &= \frac{m_W^2v^2}{f^4} c^{(8)}_{H,1},\quad C_{10}^{2h} = \frac{m_W^2v^2}{f^4} c^{(8)}_{H,2}.\\
\end{split}
\eeq
We have adopted the SILH basis~\cite{Giudice:2007fh} for the dimension-six operators:
\beq
\small
\begin{split}
&\mO_H = \frac{1}{2}\partial_\mu (H^\dagger H)  \partial^\mu (H^\dagger H), \quad {\cal O}_W =\frac{ig}{2 }\left( H^\dagger  \sigma^a \overleftrightarrow{D}^\mu H \right )D^\nu  W_{\mu \nu}^a , \\
&{\cal O}_{HW} = i g(D^\mu H)^\dagger\sigma^a(D^\nu H)W^a_{\mu\nu},
\end{split}
\label{eq:d6ops}
\eeq
and the relevant dimension-eight operators are given by
\beq
\mO^{(8)}_{H,1} = (D_\mu H^\dagger D^\mu H )^2, \quad \mO^{(8)}_{H,2} = D_\mu H^\dagger D_\nu H  D^\mu H^\dagger D^\nu H.
\eeq
The effective Lagrangian  is parameterized as
\beq
\label{eq:d6lag}
\mL^{\text{SMEFT}} = \frac{c_H}{f^2} \mO_H + \sum_{i=W,B,HW,HB} \frac{c_i}{m_{\rho}^2} \mO^{(6)}_i + \sum_{i = H,1; H,2} \frac{c_i}{f^4} \mO^{(8}_i\ .
\eeq

A few comments are in order for the operators listed in Eq.~(\ref{eq:nonL}). First we see that $C_0^{h,2h}$ have the same structures as the SM $VVh$ and $VVhh$ interactions and would only modify the coupling strength. Moreover,  the gauge bosons contained in $C_9^{h,2h}$ and  $C_{10}^{h,2h}$ also have the same structure as in the SM. 
Those operators were referred as ``genuine dim-6 Higgs operators'' \cite{Barger:2003rs,Giudice:2007fh}. We will see that for these anomalous couplings the EWA works quite well, following the dynamical structure of the SM. On the other hand, both $C_5^{h,2h}$ and $C_6^{h,2h}$ introduce new Lorentz structures with two or more derivatives on the gauge bosons, implying new underlying dynamics at a higher scale. In particular, $C_5^{h,2h}$ terms vanish for on-shell $W$ gauge bosons and $C_6^{h,2h}$ mainly contribute to transverse gauge boson processes. As we will see later on, the presence of these structures necessitates a careful examination on the interplay between EWA and full fixed-order calculations,  and provide valuable insights into the vector boson fusion (VBF) process in very high energy electroweak scatterings.

\section{Vector Boson Fusion: Anatomy of $W^+W^-\rightarrow h h$}
\label{sec:ww}

In our study we will assume the amplitudes are dominated by the SM contributions $-$ otherwise the perturbative expansion in $p^2$ would become invalid. Therefore, the leading effects of new physics beyond the SM will be probed through its interference with the SM contribution. In the following, to be self-consistent, we only keep terms linear in the Wilson coefficients $\tilde{C}^h_i$ and $\tilde{C}^{2h}_i$ in the expression.

It is informative to present a detailed analytical study of the partonic level process $W^+W^-\rightarrow h h$. We start with the helicity amplitudes  by employing the Wigner $d$-function in Section~\ref{sec:helamp}, then study the threshold behavior in Section~\ref{sec:thresh} and the high energy limits in Section~\ref{sec:HEL}.

\subsection{Helicity amplitudes}
\label{sec:helamp}
The Feynman diagrams contributing to $W^+W^-\rightarrow h h$ are shown in Fig.~\ref{fig:feyn}. The helicity amplitude  can be expressed in terms of the Wigner $d$-function $d^{J_0}_{\Delta\lambda,0}(\theta)$,
\begin{equation}
\label{eq:heldecomp}
\mathcal{M}(W^+_{\lambda}W^-_{\bar{\lambda}}\rightarrow h h ) = \widetilde{\mathcal{M}}_{\lambda\bar{\lambda}} (-1)^{\bar{\lambda}} d^{J_0}_{\Delta\lambda,0}(\theta)
\end{equation}
where $\lambda \ (\bar{\lambda})$ is the helicity of the incoming $W^+(W^-)$ boson, $\Delta\lambda=\lambda-\bar{\lambda}$ and $J_0 = |\Delta\lambda|$. We follow the convention in Ref.~\cite{Hagiwara:1986vm}. The relevant Wigner $d$-functions  are~\cite{ParticleDataGroup:2022pth},
\begin{equation}
\label{eq:dwigner}
d^0_{0,0} = 1\ ,\qquad d^1_{\pm 1,0}=\pm\frac{1}{\sqrt{2}}\sin\theta\ ,\qquad d^2_{\pm 2,0} = \sqrt{\frac{3}{8}}\sin^2\theta\ ,
\end{equation}
where $\theta$ is the polar angle of the outgoing Higgs boson in the centre-of-mass (CM) frame, with the $z$-axis defined by the incoming $W^+$ boson. Bose symmetry in the outgoing Higgs bosons requires the total amplitudes must be invariant under $\theta \to \pi-\theta$.

In Eq.~(\ref{eq:heldecomp}) $\widetilde{\mathcal{M}}_{\lambda\bar{\lambda}}$ is the reduced amplitude and can be decomposed into 4 terms corresponding to the 4 diagrams  in Fig.~\ref{fig:feyn}:
\begin{equation}
\widetilde{\mathcal{M}} = \widetilde{\mathcal{M}}^s + \widetilde{\mathcal{M}}^t + \widetilde{\mathcal{M}}^u + \widetilde{\mathcal{M}}^4 \ .
\end{equation}
In the CM frame, we have
\begin{equation}
\beta_a = \sqrt{1-\frac{4m_a^2}{s}}\ ,\qquad \gamma_a = (1-\beta_a^2)^{-1/2} = \dfrac{\sqrt s}{ 2m_a}\ , 
\ee
where $m_a$ is the mass and $\beta_a$ is the velocity of the particle $a$ in natural unit. The Mandelstam invariants $t$ and $u$ can then be written as
\be 
t=m_W^2+m_h^2 - \dfrac{s}{2}(1- \beta_W \beta_h \cos\theta)\ ,\qquad u = m_W^2+m_h^2 - \dfrac{s}{2}(1+ \beta_W \beta_h \cos\theta)\ .
\end{equation}
Since we will be studying the threshold and high-energy behaviors of the amplitudes, it will be convenient to factor the couplings and propagators out of the reduced amplitudes as the following,
\bea
\label{eq:modifiedMs}
\widetilde{\mathcal{M}}^s_{\lambda\bar{\lambda}} 
&=& \frac{g^2}{4} \frac{s}{s-m_h^2} \mathcal{A}^s_{\lambda\bar{\lambda}}
= \frac{g^2}{4} \frac{1}{1-m_h^2/s} \mathcal{A}^s_{\lambda\bar{\lambda}}\nonumber ,\\
\widetilde{\mathcal{M}}^t_{\lambda\bar{\lambda}} 
&=& \frac{g^2}{4} \frac{-s/2}{t-m_W^2} \mathcal{A}^t_{\lambda\bar{\lambda}}
= \frac{g^2}{4} \frac{1}{1-\beta_W \beta_h \cos\theta - 2m_h^2/s}\ \mathcal{A}^t_{\lambda\bar{\lambda}}\nonumber , \\
\widetilde{\mathcal{M}}^u_{\lambda\bar{\lambda}} 
&=& \frac{g^2}{4} \frac{-s/2}{u-m_W^2} \mathcal{A}^u_{\lambda\bar{\lambda}}
= \frac{g^2}{4} \frac{1}{1+\beta_W \beta_h \cos\theta - 2m_h^2/s}\ \mathcal{A}^u_{\lambda\bar{\lambda}}\nonumber , \\
\widetilde{\mathcal{M}}^4_{\lambda\bar{\lambda}} 
&=& \frac{g^2}{4} \mathcal{A}^4_{\lambda\bar{\lambda}} ,
\eea
where we have defined the ``dimensionless propagators'' by adding a factor of $s$ and $s/2$ to the $s$-channel and $t/u$-channel propagators, respectively. We compute the reduced amplitudes, including contributions from the nonlinear EFT in Eq.~(\ref{eq:nonL}) and up to terms linear in the Wilson coefficients $C_i^{h/2h}$. Using the Feynman rules  in  Appendix~\ref{appendix:B}, the full results for the helicity  amplitudes ${\cal A}^i_{\lambda\bar{\lambda}}\ (i=s, t, u, 4)$ are listed in Appendix~\ref{appendix:A}. In Table \ref{tab:helicity} we show how new physics interactions in Eq.~(\ref{eq:nonL}) contribute to the helicity amplitudes in various channels. The 3-point couplings $C_i^h$ only enter into the first three diagrams in Fig.~\ref{fig:feyn}, while the 4-point couplings $C_i^{2h}$ only enter into the last diagram. Among the three cubic couplings,  $C_{0,5}^h$ contributes to the $(\pm, \mp)$ helicity amplitudes while $C_6^h$ does not. The 4-point couplings $C_i^{2h}$, $i=5,6,9$, only shows up in the $(0,0)$ and $(\pm, \pm)$ helicities.

For later reference, we present the  helicity amplitudes in the SM explicitly here as 
\begin{alignat}{2}
\mathcal{A}^s_{0,0} =&\frac{12\lambda}{g^2}(1+\beta_W^2)\ ,  \qquad \qquad\qquad  \qquad \qquad \mathcal{A}^4_{0,0} =  2\gamma_W^2(1+\beta_W^2)\ ,\nonumber \\\
\mathcal{A}^t_{0,0} =&~ -2(1+\beta_W^2) - 2\gamma_W^2(\beta_W-\beta_h\cos\theta)^2 \ , \nonumber \\\
\mathcal{A}^u_{0,0} =&~ -2(1+\beta_W^2) - 2\gamma_W^2(\beta_W+\beta_h\cos\theta)^2\ ,\nonumber \\
\mathcal{A}^t_{\pm,0} =&~ \mathcal{A}^t_{0,\mp} =-2\gamma_W\beta_h(\beta_W-\beta_h\cos\theta) \ ,\quad
\mathcal{A}^u_{\pm,0} =~ \mathcal{A}^u_{0,\mp} =2\gamma_W\beta_h(\beta_W+\beta_h\cos\theta) \ ,\nonumber \\\
\mathcal{A}^t_{\pm,\mp} =&~ \mathcal{A}^u_{\pm,\mp}= -\frac{4}{\sqrt{6}}\beta_h^2\ , \, \qquad\qquad\qquad \qquad \mathcal{A}^s_{\pm,\pm} = -\frac{12\lambda}{g^2}\frac1{\gamma_W^{2}}\ , \nonumber \\ 
\mathcal{A}^t_{\pm,\pm} =&~\mathcal{A}^u_{\pm,\pm}= (2\gamma_W^{-2}+\beta_h^2\sin^2\theta)\ , \qquad  \qquad \mathcal{A}^4_{\pm,\pm} = -2\ . 
\end{alignat}
where $\lambda$ is the Higgs quartic coupling in the SM, and enters into the Lagrangian through $ \lambda|H|^4$. All other helicity amplitudes vanish in the SM. The polarized {\em partonic} cross section can be computed
\begin{equation} 
\hat{\sigma}_{\lambda \bar\lambda} = \frac{1}{2\beta_Ws} |\mathcal{M}_{\lambda \bar\lambda}|^2 d{\rm PS}_2\ , \quad 
\label{eq:xsec_def}
d{\rm PS}_2 = \frac{1}{2}\frac{1}{(4\pi)^2} \dfrac{\beta_h }{2} d\Omega\ .
\end{equation}
The extra factor $1/2$ is a symmetry factor for identical particles in the final state.

\begin{table}
\centering
\begin{tabular}{|c|c|c|c|c|c|c|c|c|c|c|}
\hline
Helicity & Diagram & SM & $C^h_0$ & $C^h_5$ & $C^h_6$ & $C^{2h}_0$ & $C^{2h}_5$ & $C^{2h}_6$ & $C^{2h}_9$ & $C^{2h}_{10}$ \\
\hline\hline
\multirow{4}{*}{\shortstack{$(0,0)$\\$(\pm,\pm)$}} & $s$-channel & \checkmark & \checkmark & \checkmark & \checkmark & - & - & - & - & - \\
\cline{2-11}
& $t$-channel & \checkmark & \checkmark & \checkmark & \checkmark & - & - & - & - & - \\
\cline{2-11}
& $u$-channel & \checkmark & \checkmark & \checkmark & \checkmark &- & - & - & - & - \\
\cline{2-11}
& 4-point & \checkmark & - & - & - & \checkmark & \checkmark & \checkmark & \checkmark & \checkmark \\
\hline\hline
\multirow{4}{*}{$(\pm,\mp)$} & $s$-channel & - & - & - & - & - & - & - 
& - & -\\
\cline{2-11}
& $t$-channel & \checkmark & \checkmark & \checkmark & - & - & - & - & - & - \\
\cline{2-11}
& $u$-channel & \checkmark & \checkmark & \checkmark & - & - & - & - & - & -\\
\cline{2-11}
& 4-point & - & - & - & - & - & - & -& - & \checkmark \\
\hline\hline
\multirow{4}{*}{\shortstack{$(\pm,0)$\\$(0,\pm)$}} & $s$-channel & - & - & - & - & - & - & - & - & -\\
\cline{2-11}
& $t$-channel & \checkmark & \checkmark & \checkmark & \checkmark & - & - & - & -  & - \\
\cline{2-11}
& $u$-channel & \checkmark & \checkmark & \checkmark & \checkmark & - & - & - & - & -\\
\cline{2-11}
& 4-point & - & - & - & - & - & - & - & - & \checkmark \\
\hline
\end{tabular}
\caption{\em  New physics  contributions to different helicity amplitudes in various channels. For comparison we include the SM in the table. A ``-'' denotes no contribution in that particular helicity channel.}
\label{tab:helicity}
\end{table}

\subsection{Threshold Regime}
\label{sec:thresh}

We consider the amplitudes and cross-section near the production threshold  $\sqrt s \sim 2 m_h$, which is equivalent to Taylor expand them in terms of small Higgs boson velocity $\beta_h \ll 1$. Note that in this regime, the velocity of $W$-boson  $\beta_W \approx \sqrt{1-m_W^2/m_h^2} \approx 0.77$ and the boost factor $\gamma_W \approx m_h/m_W \approx 1.55$ are all $\mathcal{O}(1)$ parameters.
This is an important kinematic region to explore because in high energy colliders most of the Higgs bosons  in the VBF channel are produced near the threshold. Note that in this case the ``dimensionless propagators'' in Eq.~(\ref{eq:modifiedMs}) approach constants 
\be
\frac{s}{s-m_h^2}\rightarrow \frac{4}{3}\ ,\qquad \frac{-s/2}{t-m_W^2}\rightarrow 2\ , \qquad \frac{-s/2}{u-m_W^2}\rightarrow 2\ .
\ee
Keeping only terms that are leading orders in $\beta_h$ in each helicity amplitude,  the amplitudes simplify to 
\begin{align}
\label{Eq:mtilde}
\widetilde{\mathcal{M}}_{0,0} =&~ ~ -g^2\left\{\frac{1}{2}(1+4\beta_W^2+2\gamma_W^2) - \frac{4\lambda}{g^2}(1+\beta_W^2) + C_0^h \left[4\beta_W^2+4\gamma_W^2-\frac{4\lambda}{g^2}(1+\beta_W^2)\right] \right. \nonumber\\
&- \frac{1}{2}C_0^{2h} \gamma_W^2(1+\beta_W^2)  + C^h_5\left[2+4\beta_W^2 - \frac{4\lambda}{g^2}(1+\beta_W^2)\right] + C^h_6\left[4\beta_W^2 + \frac{4\lambda}{g^2}\gamma_W^{-2}\right] \nonumber \\
&~\left. + C^{2h}_6 - \gamma_W^2 C^{2h}_5 (1+\beta_W^2) + \frac{1}{2}\gamma_W^4\left[ C^{2h}_9(1+\beta_W^2) + C^{2h}_{10}\beta_W^2 \right]\right\}\ , \nonumber\\
\widetilde{\mathcal{M}}_{0,\pm}=&~\widetilde{\mathcal{M}}_{0,\mp} =  2g^2\gamma_W\left[(1-2\beta_W^2)(1+2C_0^h+2C^h_5) -4C^h_6\beta_W^2 + \frac{1}{4}C^{2h}_{10}\gamma_W^2\right]\beta_h^2\cos\theta\ ,\nonumber \\
\widetilde{\mathcal{M}}_{\pm,\mp} =&~ - g^2\frac{4}{\sqrt{6}}\left( 1 + 2C_0^h + 2 C^h_5+ \frac{1}{4}C^{2h}_{10}\gamma_W^2 \right)\beta_h^2\ , \nonumber\\
\widetilde{\mathcal{M}}_{\pm,\pm} =&~ g^2\left\{\frac{3}{2}-2\beta_W^2 - \frac{4\lambda}{g^2}\gamma_W^{-2} + 4C_0^h\left(1-\frac{\lambda}{g^2}\right)\gamma_W^{-2} - \frac{1}{2}C_0^{2h} + 2C^{h}_5\left(1-2\beta_W^2 - \frac{2\lambda}{g^2}\gamma_W^{-2}\right) \right.\nonumber \\ 
&~\left. - 4C^h_6\left[\beta_W^2 - \frac{\lambda}{g^2}(1+\beta_W^2)\right]
 - C^{2h}_5 + \gamma_W^2\left[ C^{2h}_6(1+\beta_W^2) + \frac{1}{2} C^{2h}_9 \right]\right\} \ . 
\end{align}
Observe that $\widetilde{\mathcal{M}}_{0,0}$ and $\widetilde{\mathcal{M}}_{\pm,\pm}$ are constant in $\beta_h$ due to the 
$S$-wave behavior, while $\widetilde{\mathcal{M}}_{0,\pm}$, $\widetilde{\mathcal{M}}_{0,\mp}$  and $\widetilde{\mathcal{M}}_{\pm,\mp}$ are  ${\cal O}(\beta_h^2)$, owing to the higher partial waves, and are thus highly suppressed in the threshold region. Therefore we expect the production cross-section to be dominated by the $(0,0)$ and $(\pm,\pm)$ amplitudes. More explicitly, the polarized partonic  cross sections are
\begin{alignat}{2}
\hat{\sigma}_{0,0} \to&~ \frac{\pi \alpha^2}{s_W^4m_h^2 \beta_W}\frac{\beta_h}{8} \left\{\left[\frac{1}{2}(1+4\beta_W^2+2\gamma_W^2) - \frac{4\lambda}{g^2}(1+\beta_W^2)\right]^2 \right.\nonumber\\
&~ +2\left[\frac{1}{2}(1+4\beta_W^2+2\gamma_W^2) - \frac{4\lambda}{g^2}(1+\beta_W^2)\right]\left[ C_0^h \left(4\beta_W^2+4\gamma_W^2-\frac{4\lambda}{g^2}(1+\beta_W^2)\right)  \right.\nonumber\\ 
&~ - \frac{1}{2}C_0^{2h}\gamma_W^2(1+\beta_W^2) + C^h_5\left(2+4\beta_W^2 - \frac{4\lambda}{g^2}(1+\beta_W^2)\right) + C^h_6\left(4\beta_W^2 + \frac{4\lambda}{g^2}\gamma_W^{-2}\right)\nonumber \\
&~ \left.\left. + C^{2h}_6 - \gamma_W^2 C^{2h}_5 (1+\beta_W^2) + \frac{1}{2}\gamma_W^4\left( C^{2h}_9(1+\beta_W^2) + C^{2h}_{10}\beta_W^2 \right)\right]\right\} \ , \nonumber\\
\hat{\sigma}_{\pm,0}=&~\hat{\sigma}_{0,\pm} \to
\frac{\pi \alpha^2}{s_W^4m_h^2 \beta_W}{\beta_h^5\over 30} {m_h^2\over m_W^2}  \left\{(1-2\beta_W^2)^2 + 2(1-2\beta_W^2) \left[2(1-2\beta_W^2)(C_0^h+C^h_5) \vphantom{\frac{1}{2}}\right.\right.\nonumber \\
&~\left.\left. -4C^h_6\beta_W^2 + \frac{1}{4}C^{2h}_{10}\gamma_W^2\right]\right\}, \nonumber\\
\hat{\sigma}_{\pm,\mp} \to&~ \frac{\pi \alpha^2}{s_W^4m_h^2 \beta_W}\frac{\beta_h^5}{15}\left[1 + 2\left(2C_0^h + 2 C^h_5+ \frac{1}{4}C^{2h}_{10}\gamma_W^2\right)\right] ,\nonumber \\
\hat{\sigma}_{\pm,\pm} \to&~ \frac{\pi\alpha^2}{s_W^4m_h^2 \beta_W}\frac{\beta_h}{8}\left\{\left(\frac{3}{2}-2\beta_W^2 - \frac{4\lambda}{g^2}\gamma_W^{-2}\right)^2 + 2\left(\frac{3}{2}-2\beta_W^2 - \frac{4\lambda}{g^2}\gamma_W^{-2}\right)\left[ 4C_0^h\left(1-\frac{\lambda}{g^2}\right)\gamma_W^{-2}\right.\right.\nonumber \\
&~ - \frac{1}{2}C_0^{2h} + 2C^{h}_5\left(1-2\beta_W^2 - \frac{2\lambda}{g^2}\gamma_W^{-2}\right) 
 - 4C^h_6\left(\beta_W^2 - \frac{\lambda}{g^2}(1+\beta_W^2)\right) \nonumber \\
&~\left.\left. - C^{2h}_5 + \gamma_W^2\left( C^{2h}_6(1+\beta_W^2) + \frac{1}{2} C^{2h}_9 \right)\right]\right\} \ ,
\end{alignat}
where $\alpha=g^2s_W^2/4\pi$.

To gain some intuition on the relative contributions from different operators, 
it is instructive to plug in the explicit numbers near threshold. In this regime $\widetilde{\mathcal{M}}_{0,\pm}$ and $\widetilde{\mathcal{M}}_{\pm,\mp}$ are all of order $\beta_h^2$, according to Eqs.~(\ref{Eq:mtilde}), and hence can be neglected. The remaining helicity amplitudes are $\widetilde{\mathcal{M}}_{0,0}$ and $\widetilde{\mathcal{M}}_{\pm,\pm}$,
\bea
\label{eq:helM00}
\widetilde{\mathcal{M}}_{0,0} &=& -g^2\left[2.17+\frac{v^2}{f^2}\left(10.09 \widetilde{C}_0^h-1.92 \widetilde{C}_0^{2h} +2.43 \widetilde{C}_5^h+2.84 \widetilde{C}_6^h\right.\right.\nonumber\\
&&\left.\left.\qquad\qquad\qquad\qquad -3.83 \widetilde{C}_5^{2h}+\widetilde{C}_6^{2h}\right) +\frac{v^4}{f^4}\left(4.63 \widetilde{C}_9^{2h}+ 1.71 \widetilde{C}_{10}^{2h}\right)\right]\ , \nonumber\\
\widetilde{\mathcal{M}}_{\pm,\pm}&=& -g^2\left[0.17+\frac{v^2}{f^2}\left(-1.16 \widetilde{C}_0^h+0.5 \widetilde{C}_0^{2h}+0.84 \widetilde{C}_5^h+0.43 \widetilde{C}_6^h\right.\right.\nonumber\\
&&\left.\left.\qquad\qquad\qquad\qquad+\widetilde{C}_5^{2h}-3.83 \widetilde{C}_6^{2h}\right)-1.21 \frac{v^4}{f^4}\widetilde{C}_9^{2h} \right]\ ,
\eea
where we have indicated the natural power counting of the Wilson coefficients by way of Eq.~(\ref{Eq:powerCC}). All the hatted quantities are expected to be order unity.
The dominant partonic cross sections arise from the longitudinal and transverse polarizations:
\begin{alignat}{2}
\label{eq:sigmaLL}
\hat{\sigma}_{LL}\to&~ \hat{\sigma}_{0,0} = \hat{\sigma}_{0}  \beta_h \left[0.59+\frac{v^2}{f^2}\left( 5.48\widetilde{C}_0^h - 1.04\widetilde{C}_0^{2h}+ 1.32\widetilde{C}^h_5 + 1.54\widetilde{C}^h_6 - 2.08\widetilde{C}^{2h}_5  \right.\right.\nonumber\\
  &\qquad \qquad \qquad\qquad\qquad\qquad \left.\left.+ 0.54 \widetilde{C}^{2h}_6\right) + \frac{v^4}{f^4}\left(2.51 \widetilde{C}^{2h}_9 + 0.93 \widetilde{C}^{2h}_{10}\right)\right] \ , \nonumber \\  
\hat{\sigma}_{TT} \to&~ \frac{1}{2}\hat{\sigma}_{+, +} = \hat{\sigma}_{0}  \beta_h \left[ 0.002-\frac{v^2}{f^2}\left(0.025\widetilde{C}_0^h - 0.011\widetilde{C}_0^{2h} + 0.018\widetilde{C}^h_5 - 0.009\widetilde{C}^h_6 \right.\right.\nonumber\\
& \qquad \qquad \qquad\qquad\qquad\qquad \left.\left.+ 0.022 \widetilde{C}^{2h}_5 + 0.082 \widetilde{C}^{2h}_6\right) - 0.026\frac{v^4}{f^4} \widetilde{C}^{2h}_9\right]\ ,
\end{alignat}
where $\sigma_0=\pi \alpha^2 /s_W^2m_h^2\beta_W$ and terms neglected are of ${\cal O}(\beta_h^2)$. 

Note that, near the threshold region, the longitudinal cross section is about two orders of magnitude larger than the transverse cross section as a result of total angular momentum conservation. This will have important implications when we discuss  the effective $W$  approximation and the electroweak parton distribution functions~\cite{Han:2020uid,Garosi:2023bvq}. It is also interesting to see in Eq.~(\ref{eq:sigmaLL})  that $\widetilde{C}_0^h$ term has an anomalously large  coefficient, which appears to be a numerical accident. This feature is already present in the helicity amplitude  in Eq.~(\ref{eq:helM00}), and gets further enhanced after the phase space integration in the cross-section.


\subsection{High Energy Limit}
\label{sec:HEL}

Next we consider the behavior of the amplitudes in the very high energy limit,
\be
 s \gg m_W^2,\ m_h^2\ ; \qquad  \beta_W,\beta_h\to 1\ ; \qquad \gamma_W,\gamma_h \to \infty\ .
\ee
In this limit, the longitudinal polarization of the vector boson scales with its momentum, leading to the sensitive probe of the high-energy behavior as already discussed in the introduction.
Two important features become prominent in this limit: 
\begin{itemize}
\item the particles involved in the scattering amplitudes become effectively massless, and the $t$- and $u$- channels exhibit the collinear singularity which enhances the scattering in the forward and backward directions, respectively. This can be seen directly from the behaviors of the dimensionless propagators in Eqs.~(\ref{eq:modifiedMs}), which in the high energy limit yield 
\begin{equation}
\frac{s}{s-m_h^2}\rightarrow 1\ , \ \ \qquad \frac{-s/2}{t-m_W^2}\rightarrow \frac{1}{1-\cos\theta}\ ,\ \  \qquad \frac{-s/2}{u-m_W^2}\rightarrow \frac{1}{1+\cos\theta}\ .
\end{equation}
\item the amplitude now exhibits energy growing behavior because the $VVh$ and $VVhh$ vertices are modified from the SM expectations and the cancellation of energy-growing terms in the amplitudes becomes incomplete. Unitarity will be violated at a certain energy and new physics is expected at that particular scale.
\end{itemize}
These considerations motivate the following decomposition of the amplitudes according to their $\sqrt{s}$ dependence~\cite{Contino:2010mh}:
\begin{equation}
\label{eq:heldecom}
\aligned
\mathcal{M} =~& \frac{-s/2}{t-m_W^2}\mathcal{A}_t + \frac{-s/2}{u-m_W^2}\mathcal{A}_u + \mathcal{A}_{\rm reg} +\frac{\sqrt{s}}{2m_W} \mathcal{A}^{(1)}\\
~&+ \left(\frac{\sqrt{s}}{2m_W}\right)^2 \mathcal{A}^{(2)} + \left(\frac{\sqrt{s}}{2m_W}\right)^3 \mathcal{A}^{(3)} + \left(\frac{\sqrt{s}}{2m_W}\right)^4 \mathcal{A}^{(4)}\ .
\endaligned
\end{equation}
Coefficients in the above equation  are given in Table \ref{tab:mhigheL}, where the contribution from each helicity amplitude is made explicit. We have also collected the leading energy dependence from each operator in the nonlinear Lagrangian in Eq.~(\ref{eq:nonL}), including the SM contribution. Notice that $\mathcal{M}_{\rm reg}$, which is constant in $\sqrt{s}$, contains contributions from both the $s$-channel  and the 4-pt diagrams in Fig.~\ref{fig:feyn}. In Table \ref{tab:leadingS}, we indicate the leading $s$ behavior for the helicity amplitudes for different anomalous couplings.

\begin{table}[t]
\small
\centering
\begin{tabular}{|c||c|c|c|c|}
\hline
Helicity   &  $(0,0)$   &  $(\pm,\pm)$   &  $(\pm,\mp)$   & $(\pm,0)$  \\
\hline\hline
$\mathcal{A}_t$   &  $-g^2(1+2C_0^h+2C_5^h)$   &  0   &  0   &  0  \\
\hline
$\mathcal{A}_u$   &  $ -g^2(1+2C_0^h+2C_5^h)$   &  0   &  0   &  0  \\
\hline
$\mathcal{A}_{\rm reg}$   &  $\begin{array}{c} 2\lambda +g^2/2\\- g^2\left(2C^h_6+C^{2h}_6\right)\end{array}$   & $ \begin{array}{c} g^2(-C_0^h+{C_0^{2h}}/{2}+C^h_5\\+C^{2h}_5)+ (2g^2-6\lambda)C^h_6 \end{array} $   &  $\begin{array}{c} g^2\left(\frac12+C_0^h+C^h_5\right)\end{array}$   &  0 \\
\hline
$\mathcal{A}^{(1)}$   &  0   &  \phantom{$\frac{g^2}2$}0   &  0   &  $\mp\sqrt{2}g^2C^h_6 \cot\theta$  \\
\hline
$\mathcal{A}^{(2)}$   &  $\begin{array}{c} g^2(-2C_0^h+C_0^{2h}\\ +2C^h_5+2C^{2h}_5)\end{array}$   &  $\begin{array}{c}-g^2\left(2C^{2h}_6 + C^{2h}_9\right.\\ \left. +\frac{1}{4} C^{2h}_{10}\ \sin^2\theta\right)\end{array}$   &  $\frac{g^2}{4} C^{2h}_{10} \ \sin^2\theta$   &  0 \\
\hline
$\mathcal{A}^{(3)}$   &  0   &  0   &  0   &  $\pm\frac{g^2}{4\sqrt{2}}C^{2h}_{10}\ \sin2\theta$ \\
\hline
$\mathcal{A}^{(4)}$  &  $\begin{array}{c}-g^2\left(2C^{2h}_9\right.\\ \left.+ \frac{1+\cos^2\theta}{2}C^{2h}_{10}\right)\end{array}$  &  0  &  0  &  0 \\
\hline
\end{tabular}
\caption{\em Helicity amplitudes in the high-energy limit, using the decomposition in Eq.~(\ref{eq:heldecom}). \label{tab:mhigheL}}
\end{table} 

\begin{table}[t]
\renewcommand{\arraystretch}{1.5}
\centering
\begin{tabular}{|c|c|c|c|c|c|c|c|c|c|}
\hline
Helicity & SM & $C_0^h$ & $C_0^{2h}$ & $C^h_5$ & $C^h_6$ & $C^{2h}_5$ & $C^{2h}_6$ & $C^{2h}_9$ & $C^{2h}_{10}$ \\
\hline\hline
$(0,0)$ &  $s^0$ & $s$ & $s$ & $s$ & $s^0$ & $s$ & $s^0$ & $s^2$ & $s^2$\\ 
\hline
$(\pm,\pm)$ & $s^{-1}$ & $s^0$ & $s^0$ & $s^0$ & $s^0$ & $s^0$ & $s$ & $s$ & $s$\\ 
\hline
$(\pm,0)$ & $1/\sqrt{s}$ & $1/\sqrt{s}$ & - & $1/\sqrt{s}$ & $\sqrt{s}$ & - & - & - & $s^{3/2}$\\ 
\hline
$(\pm,\mp)$ & $s^0$ & $s^0$ & - & $s^0$ & - & - & - & - & $s$\\ 
\hline
\end{tabular}
\caption{\em The leading energy dependence of the the amplitudes in the central region $\theta \sim \pi/2$ in the high-energy $\sqrt{s}\to \infty$ limit. $s^0$ means the amplitude is independent of $s$, while "-" implies the particular operator does not contribute to the helicity amplitude. \label{tab:leadingS}}

\end{table}

Let's first consider the SM amplitude in the high energy limit,
\begin{equation}
\label{eq:smhighE}
\mathcal{M}^{(\text{SM})} = \frac{g^2}2\left(1- \frac{2}{1-\cos\theta} -  \frac{2}{1+\cos\theta} \right) + 2\lambda \ ,
\ee
where we have put back the $s$-channel propagator to write the expression in a more illuminating form. In the expressions above, only the leading terms in $s$ are kept. In particular, after regulated by masses, the divergences in the forward/backward region lead to 
\begin{equation}
\label{eq:integral}
\int d\cos\theta \frac{1}{1\mp\cos\theta}  \rightarrow \ln\frac{s}{m_W^2},\quad 
\int d\cos\theta \left(\frac{1}{1\mp\cos\theta}\right)^2 \rightarrow \frac{s}{2m_W^2}. 
\end{equation}
While growing rather slowly, the logarithmic terms are not significantly larger than constant terms at realistic energies for future colliders. Therefore, we also keep the constant terms next to the logarithmic terms in the expressions.  Taking the high energy limit of the helicity amplitudes involving the anomalous couplings, after the integration over the angle, the partonic helicity cross sections are
\begin{align}
\label{eq:partxshe}
\hat{\sigma}_{0,0} \approx&~ \frac{\pi\alpha^2}{4s_W^4m_W^2} \left[ 1 + \left(2(C^h_0 -C^{2h}_0/2- C^{h}_5 - C^{2h}_5) + (2C^h_6 + C^{2h}_6) \frac{4m_W^2}{s} + 2 C^{2h}_9 \frac{s}{4m_W^2} \right)\right.\nonumber \\ 
&~\times\left(\ln\frac{s}{m_W^2} - \frac{1}{2}-\frac{m_h^2}{4m_W^2}\right) + 4(C^h_0+C^h_5) - C^h_6\left(2 - \frac{m_h^2}{m_W^2}\right)\frac{4m_W^2}{s} \nonumber\\
&~\left. + C^{2h}_{10} \frac{s}{4m_W^2}\left(\ln\frac{s}{m_W^2} - \frac{4}{3} - -\frac{m_h^2}{6m_W^2} \right) \right] \nonumber\\
\hat{\sigma}_{\pm,0},\hat{\sigma}_{0,\pm} \approx&~
\frac{\pi\alpha^2}{8s_W^4s}\left(1-\frac{m_h^2}{2m_W^2}\right) \left[\left( (1 +4C^h_0 + 4C^{h}_5)\left(1-\frac{m_h^2}{2m_W^2}\right)\frac{4m_W^2}{s} - 4C^h_6\right)\left(\ln\frac{s}{m_W^2}-3\right)\right. \nonumber \\
&~ \left. +\frac{2}{3} C^{2h}_{10}\frac{s}{4m_W^2}\right], \nonumber\\
\hat{\sigma}_{\pm,\mp} \approx&~ \frac{\pi\alpha^2}{8s_W^4s}\left[1 +4C^h_0 + 4C^h_5 + \frac{2}{3} C^{2h}_{10} \frac{s}{4m_W^2}\right], \nonumber\\
\hat{\sigma}_{\pm,\pm} \approx&~ \frac{\pi\alpha^2m_W^2}{4s_W^4s^2} \left[ 1 + 4 \left(C^h_0-C^h_5 - \left(2 - \frac{3m_h^2}{4m_W^2}\right)C^h_6 - C^{2h}_5 + \left(2C^{2h}_6 + C^{2h}_9\right)\frac{s}{4m_W^2}\right)\right.\nonumber \\
&~ \times \left(\ln\frac{s}{m_W^2} + 2 - \frac{5m_h^2}{2m_W^2}\right) + 4(C^h_0+C^h_5) + 4C^h_6\left(2-\frac{m_h^2}{m_W^2}\right) \nonumber\\
&~ \left. + 5C^{2h}_{10}\left(\frac{2}{3} - \frac{m_h^2}{3m_W^2}\right)\frac{s}{4m_W^2}\right]. 
\end{align}

\begin{figure}[tb]
\centering
\includegraphics[width=0.6\textwidth]{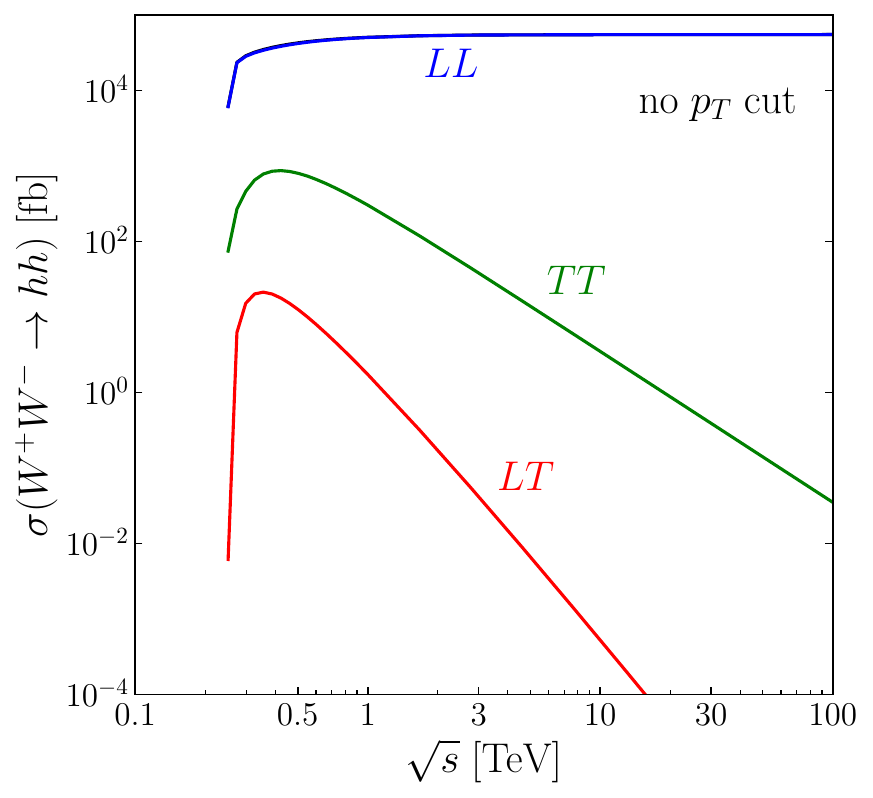}
\caption {\em Partonic cross sections of $W^+W^-
\rightarrow hh$ with longitudinal-longitudinal $(LL)$, transverse-longitudinal $(TL)$, and transverse-transverse $(TT)$ polarized initial state, respectively.}
\label{fig:partonic_xsec}
\end{figure}

In Fig.~\ref{fig:partonic_xsec}, 
we present the SM partonic cross sections as a function of the partonic center-of-mass energy for the process $W^+W^-\rightarrow hh$, using the amplitudes given in Eq.~(\ref{eq:modifiedMs}) and further detailed in Appendix~\ref{appendix:A}  with different polarized initial $W^+W^-$ states, $LL,\ TL$ and $TT$. Excluding the anomalous Higgs couplings, the constant behavior of the $LL$ cross section seen in  Fig.~\ref{fig:partonic_xsec} is due to the $t$-channel singularity as discussed in the previous section. The $TL$ component falls quickly as $(1/s)  \ln(s/m_W^2)$, and the $TT$ contribution falls even faster and scales like $1/s$, as naively expected. It is thus important to determine the high energy behavior of the cross sections in the hope to identify the BSM new physics.


\section{Beyond VBF at High Energies}

In the previous section, we have studied in detail the  helicity amplitudes and  partonic cross section for the $W^+ W^- \rightarrow h h $. To apply our results to practical applications, it is important to establish the extent to which the commonly adopted approach, the effective $W$-approximation (EWA), is valid when dealing with higher dimensional operators. For illustration, we analyze the double Higgs production at a multi-TeV muon collider by comparing the EWA method and the full LO matrix element calculations. We will compare the two methods in detail for the  SM case in Section~\ref{sec:ewasm}, while in Section~\ref{sec:ewabsm} we will also include the contribution from the anomalous couplings. {It is important to note that, although we adopt the simplest formalism in EWA for the sake of illustration, our results are equally applicable to the general  partonic framework in terms of the electroweak (EW) parton distribution functions (PDFs) in the SM \cite{Chen:2016wkt,Bauer:2017isx,Han:2020uid,Han:2021kes,Ruiz:2021tdt}.} 

Particle splitting is the dominant phenomena in the EW sector of the SM at very high energies $s \gg m_W^2, m_h^2$. The EWA approach is the generalization of the effective photon approximation to the (nearly massless) EW sector, and the approximate formulation for the partonic picture of the high energy colliding beams. 
In the high energy scattering, we expect that the short-distance physics will be factorized from the long-distance (collinear or soft) behavior.  Similar to the QCD case, the electroweak scattering total cross section can be written as  the hard-scattering sub-processes convolved with the parton distribution function of the $W,Z$ gauge bosons. For the Higgs pair production in $\mu^+\mu^-$ collisions, we write 
\begin{equation}
\label{eq:Wpdf}
\sigma\left[\mu^+\mu^- \rightarrow h h \nu_\mu \bar{\nu}_\mu\right] = \int^1_{\frac{4m_h^2}{S}} d\tau \sum_{h_1,h_2}\frac{dL_{h_1h_2}}{d\tau} \hat{\sigma}\left[W^+_{h_1}W^-_{h_2}\rightarrow h h\right](\tau S).
\end{equation}
with the polarized parton  luminosities defined as 
\begin{align}
\frac{dL_{h_1 h_2}}{d\tau} =& \int^1_\tau f_{\mu^+/W^+_{h_1}}(x,Q^2)f_{\mu^-/W^-_{h_2}}\left(\frac{\tau}{x},Q^2\right)\frac{dx}{x} 
\end{align}
Here $S$ is the center-of-mass energy squared of the muon pairs and $\tau = s_{hh}/S$ is the ratio between the invariant mass of the two Higgs bosons and invariant mass of the two muons. 
Much progress has been made recently in developing the EW PDFs
\cite{Chen:2016wkt,Bauer:2017isx,Han:2020uid,Han:2021kes,Ruiz:2021tdt}. 
For simplicity, we adopt the leading order gauge boson PDFs for un-polarized leptons, and the transverse ($\pm$) and the longitudinal ($L$) vector bosons are given by~\cite{Dawson:1984gx,Chen:2016wkt,Ruiz:2021tdt} 
\begin{align}
f_{\mu^-/V_+}(x,Q^2) =& f_{\mu^+/V_-}(x,Q^2) = \frac{1}{16\pi^2 x}(C_L^2(1-x)^2+C_R^2)\ln\frac{Q^2}{m_V^2},\\
f_{\mu^-/V_-}(x,Q^2) =& f_{\mu^+/V_+}(x,Q^2) = \frac{1}{16\pi^2 x}(C_L^2+C_R^2(1-x)^2)\ln\frac{Q^2}{m_V^2},\\
f_{\mu^-/V_L}(x,Q^2) =& f_{\mu^+/V_L}(x,Q^2) = \frac{C_L^2+C_R^2}{8\pi^2 }\frac{1-x}{x}.
\end{align}
where $Q$ is the factorization scale of PDF and for $W$-boson:
\beq
C_V = - C_A = \frac{g}{2\sqrt{2}},\qquad 
C_L = \frac{g}{\sqrt{2}},\qquad C_R = 0,
\eeq
while for $Z$-boson:
\beq
C_V = \frac{g}{\cos\theta_W}\left( \frac12 T^{3L} - Q \sin^2\theta_W\right),\quad 
C_A = - \frac{g}{\cos\theta_W} \frac12 T^{3L}. 
\eeq
\beq
C_L = \frac{g}{\cos\theta_W}\left(  T^{3L} - Q \sin^2\theta_W\right), \quad 
C_R = - \frac{g}{\cos\theta_W} Q \sin^2\theta_W. 
\eeq
In contrast to the photons and gluons, we have the longitudinal splitting function. It comes from the power correction $m_W^2/p^2_T$ and can be much larger than the Yukawa coupling contributions of the Goldstone boson to light fermions. 
It is not enhanced by the large logarithmic factors and exhibits an  ``ultra collinear'' behavior \cite{Chen:2016wkt}.
Due to the larger coupling for the $W$-boson than for the $Z$-boson, we expect the cross section for the double Higgs production to be dominated by  $W^+W^-$ fusion.

\begin{figure}[tb]
\centering
\includegraphics[width=.48\textwidth]{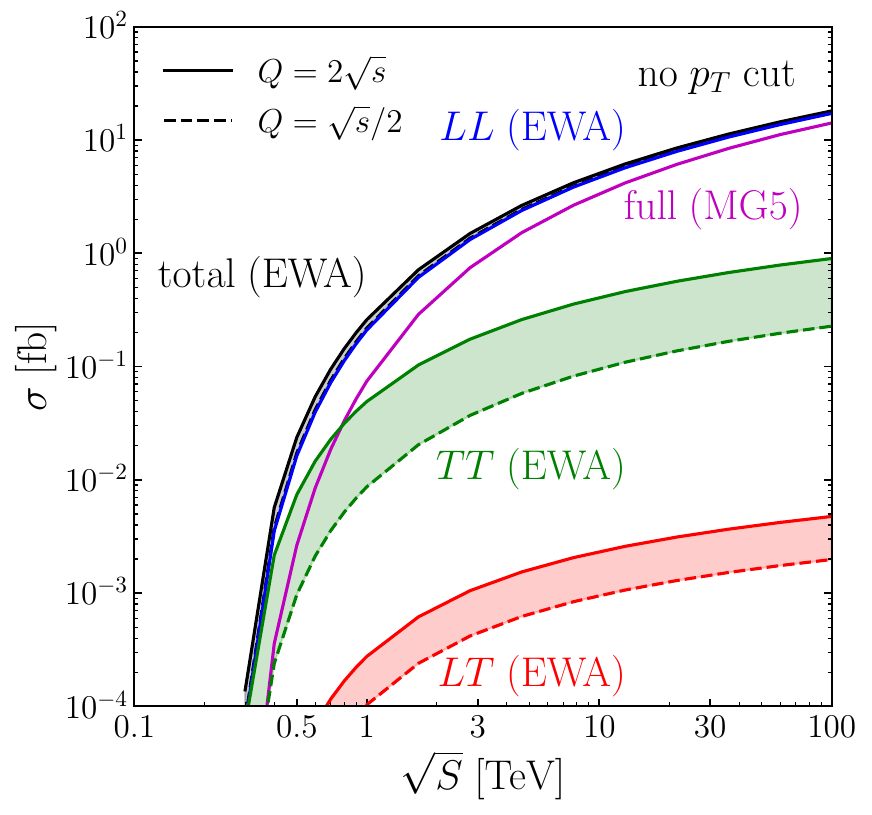}
\includegraphics[width=.48\textwidth]{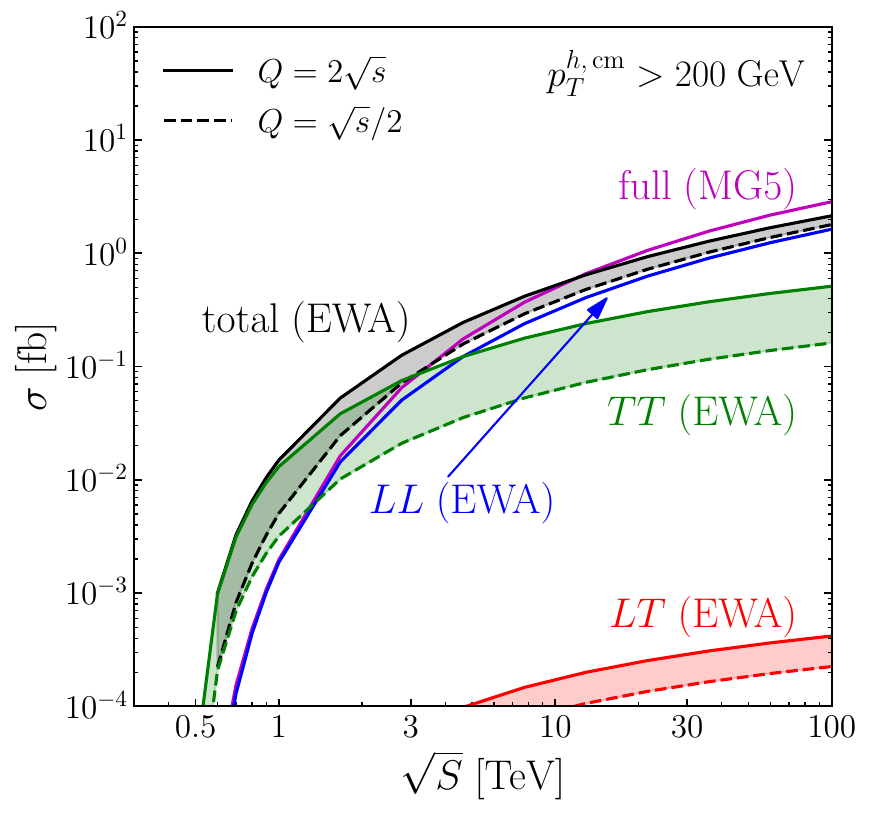}
\caption{\em Left: contributions from different helicity configurations to the SM cross sections of $\mu^+\mu^- \rightarrow h h \nu_\mu \bar{\nu}_\mu$ as functions of the $\mu^+\mu^-$ collision energy $\sqrt{S}$, using EWA. The blue, green, and red lines represent longitudinal-longitudinal $(LL)$, transverse-transverse $(TT)$, and longitudinal-transverse $(LT)$ helicities, respectively. The black lines represent the sum of all helicities, and the purple line is the full LO matrix element calculation by MadGraph5~\cite{Alwall:2014hca}. No $p_T$ cut applied on the Higgs pair. The bands correspond to varying the $W$ PDF scale from $Q = \sqrt{s}/2$ (dashed) to $Q=2\sqrt{s}$ (solid). Right: same as the left plot but with cut on the transverse momentum of the Higgs bosons  $p_T^{h, \rm cm} > 200$ GeV.}
\label{fig:smxsec_1}
\end{figure}

\subsection{SM case}
\label{sec:ewasm}
In this subsection, we will compare EWA approach with the MadGraph5 leading order (LO) calculations \cite{Alwall:2014hca} in the SM case~\cite{Han:2020uid,Ruiz:2021tdt}. The results obtained by the two methods are shown in Fig.~\ref{fig:smxsec_1}, where different $W^+W^-$ helicity contributions for the EWA calculations as in Eq.~(\ref{eq:Wpdf}) are also presented.  We have shown two  cross sections: one without any kinematic cut and one with minimal cut of 200 GeV  on the transverse momenta of the Higgs bosons in the Higgs boson pair rest frame ($p_T^{h,\rm cm} > 200 $ GeV). In EWA, this is the partonic CM frame. 
Moreover, we present the EWA calculation by varying the $W$ PDF  factorization scale from  $\sqrt{s}/2$ to  $2 \sqrt{s}$ as an estimate of uncertainties due to higher order corrections. 
{In general, the resummation of the higher order logarithms may lead to an enhancement of the production cross section beyond the tree-level calculations, as already established in QCD calculations.} 

First we note that, EWA tends to yield a larger cross section than that from a leading order result by MG5 without any kinematical cuts. The main cause is due to the threshold effects at the order of $m_W^2/s,\ m_W^2/t$. When we apply the hardness cut $p_T^{h,\rm cm} > 200$ GeV to improve its validity $s,|t|,|u| \gg m_W^2$, the two methods agree better, especially within the uncertainty bands determined by varying the PDF factorization scales. This behavior is expected from the theoretical study of the literature~\cite{Dawson:1984gx,Kunszt:1987tk,Borel:2012by}. Secondly, as expected from the partonic cross section formula, the $LL$ mode dominates over all the energy regime due to the $t$-channel singularity, although in the real experiment, we always have  a finite range cover of the rapidity. In fact, we do see from the right panel of Fig.~\ref{fig:smxsec_1} that after the $p_T^{h,\rm cm}$ cut, the dominance of the $LL$ mode becomes weaker and  at low CM energy of the muon pairs, $TT$ modes start to become dominant. The contribution from the $LT$ modes quickly becomes irrelevant after the hardness cut, as can be  understood from the Goldstone equivalence theorem. 
Finally, as discussed earlier, the threshold behavior in terms of the power of the speed of the Higgs boson $\beta_h = \sqrt{1 -{m_h^2}/{E_h^2}}$ governs the cross section increase over the CM energy.

\subsection{EWA in the presence of anomalous couplings}
\label{sec:ewabsm}

\begin{figure}[tb]
\centering
\includegraphics[width=.48\textwidth]{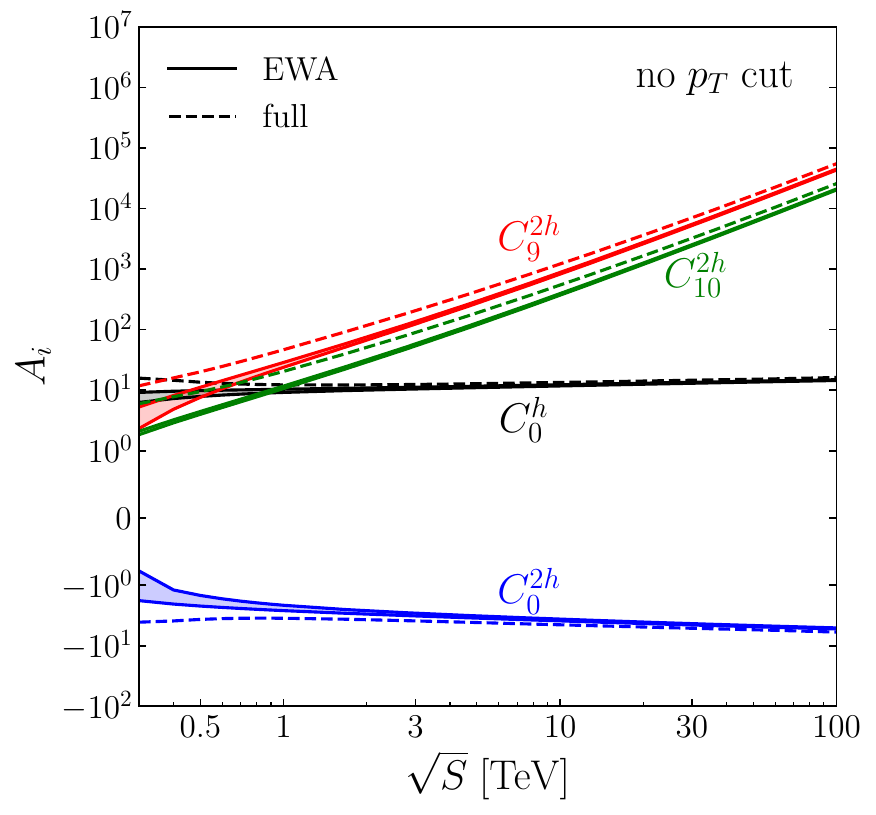}
\includegraphics[width=.48\textwidth]{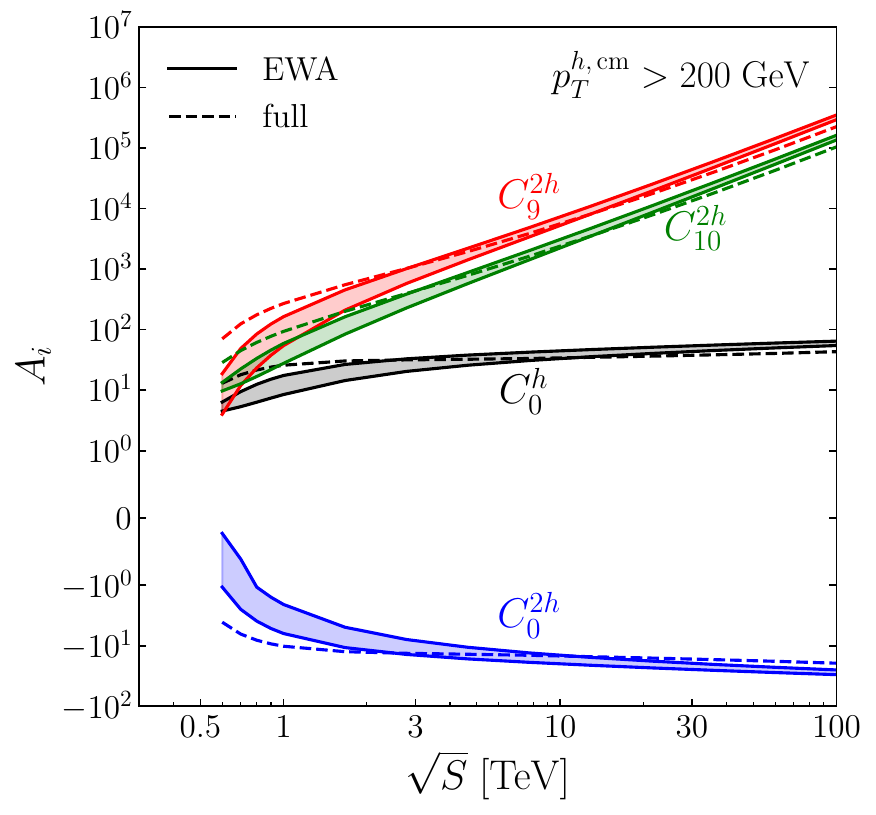}
\caption{\em The linear dependence $A_i$ on the anomalous couplings, as defined $\sigma (\mu^+ \mu^- \rightarrow h h \nu \bar{\nu}) = \sigma_{\rm SM} \left(1 + A_i C_i + \cdots \right)$. The solids lines are computed using EWA, and the dashed lines come from the full matrix element calculation by MadGraph5. The bands correspond to varying the $W$ PDF scale from $Q = \sqrt{s}/2$ to $Q=2\sqrt{s}$. Left panel: No $p_T$ cut applied on the Higgs pair. Right panel: same as the left plots but with cut on the transverse momentum of the Higgs bosons  $p_T^{h, \rm cm} > 200$ GeV.}
\label{fig:coefs_1}
\end{figure}
\begin{figure}[tb]
\centering
\includegraphics[width=.48\textwidth]{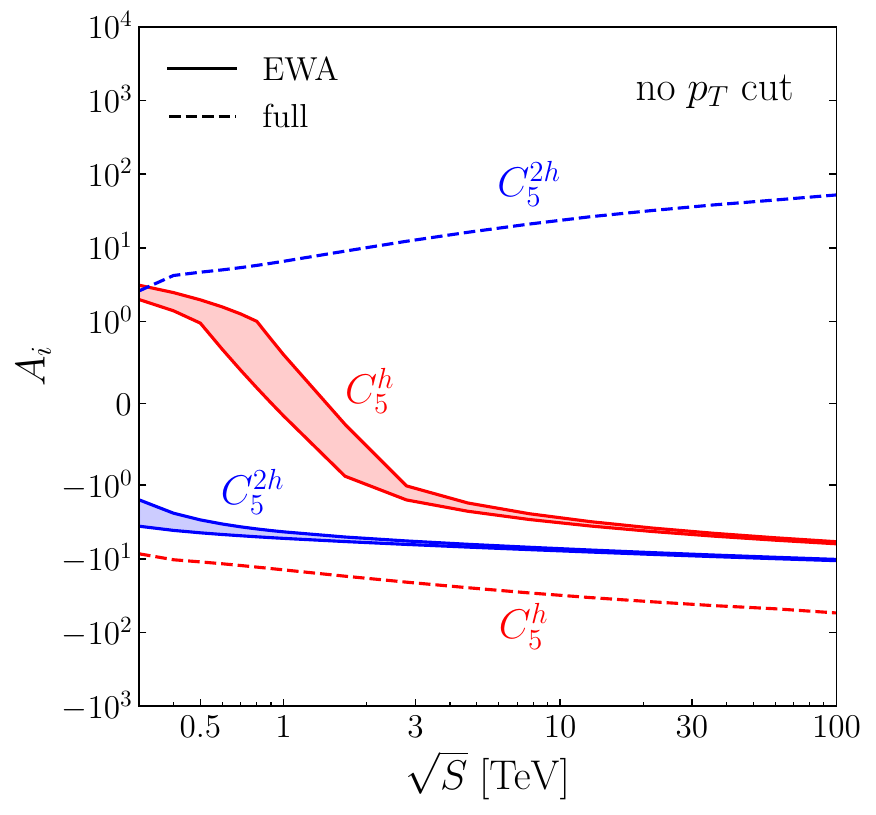}
\includegraphics[width=.48\textwidth]{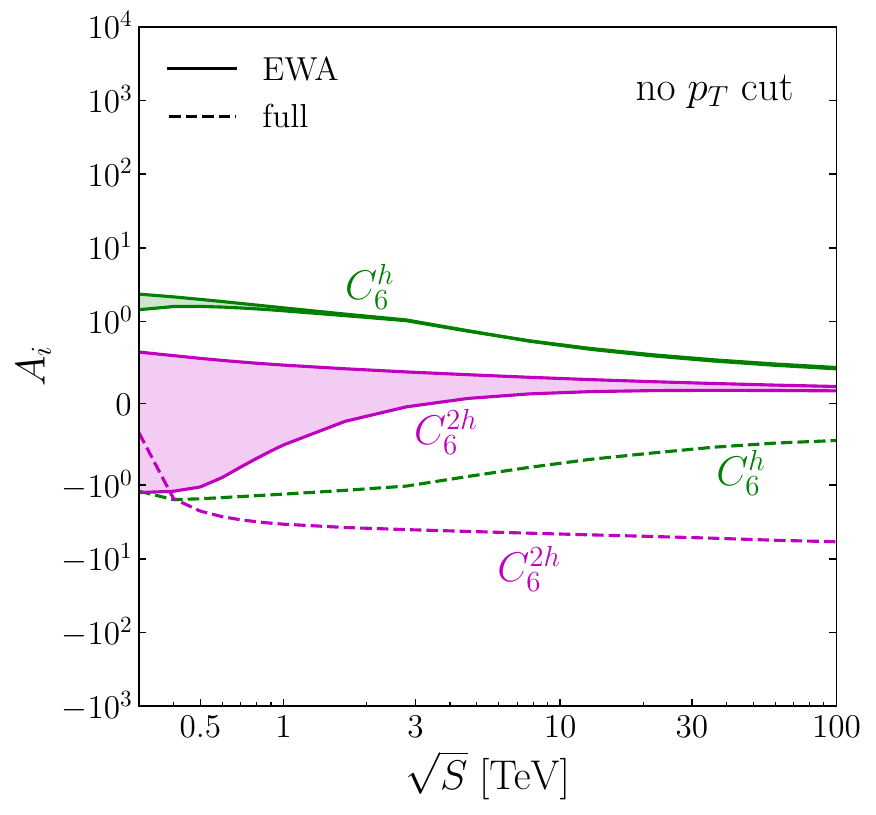}
\caption{\em The linear coefficients $A_i$ for the anomalous couplings $C_5^{h,2h}$ and $C_6^{h,2h}$ without $p_T$ cuts and see the caption of  Fig.~\ref{fig:coefs_1} for detailed description. }
\label{fig:coefs_2}
\end{figure}
\begin{figure}[tb]
\centering
\includegraphics[width=.48\textwidth]{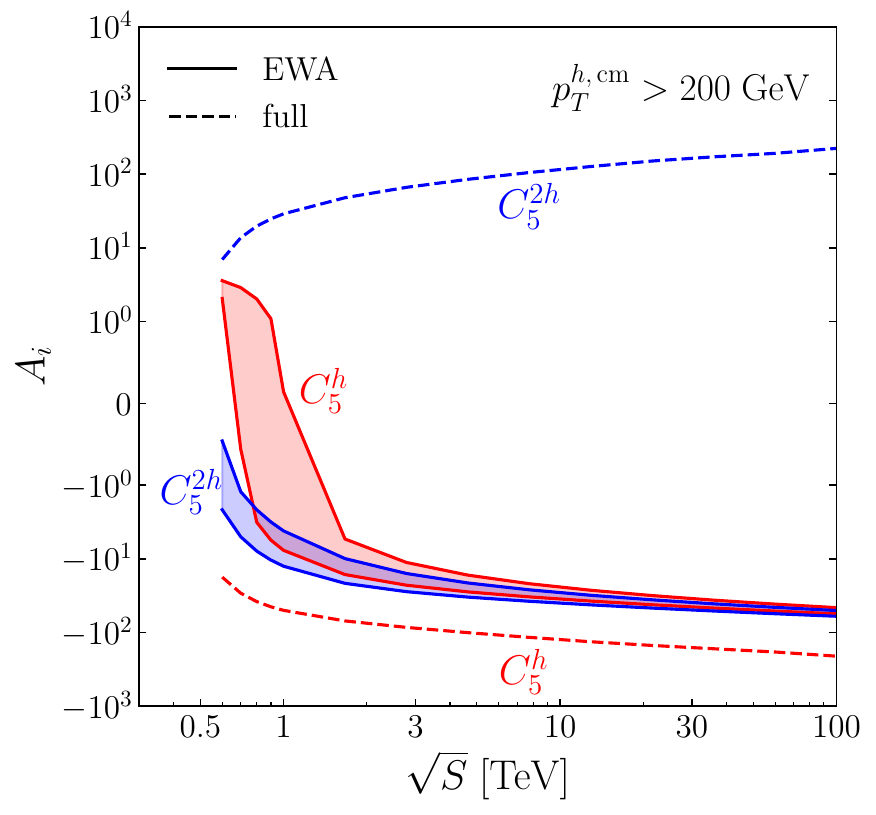}
\includegraphics[width=.48\textwidth]{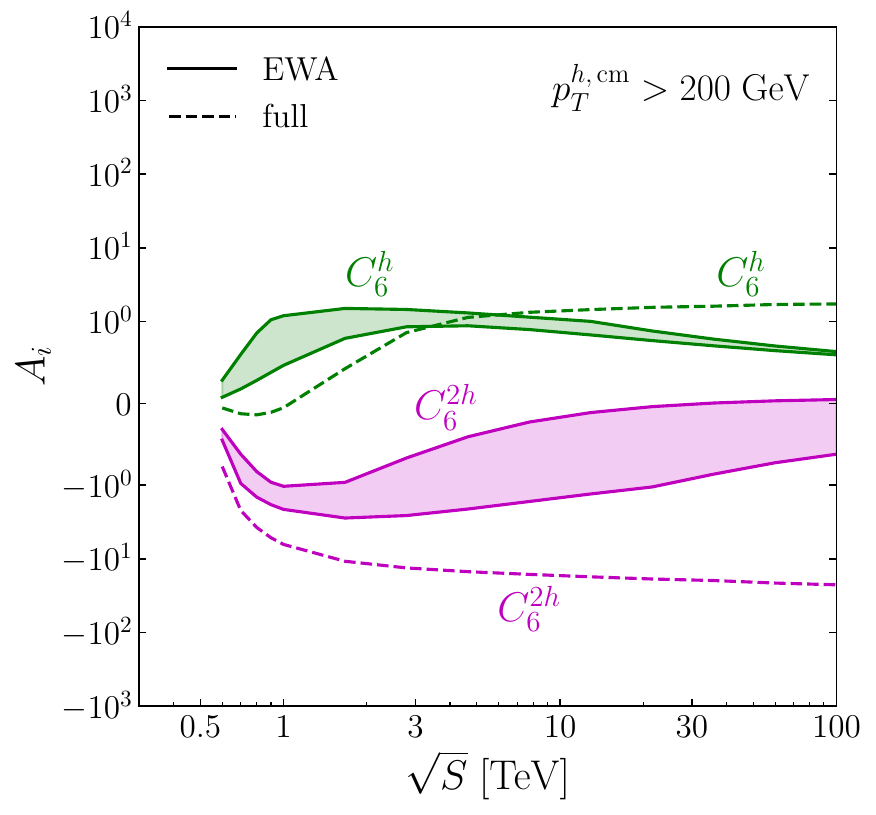}
\caption{\em The linear coefficients $A_i$ for the anomalous couplings $C_5^{h,2h}$ and $C_6^{h,2h}$ with the cut $p_T^{h,cm} > 200$ GeV and see the caption of  Fig.~\ref{fig:coefs_1} for detailed description. }
\label{fig:coefs_3}
\end{figure}

We now compare the EWA results with the full LO calculation by switching on the anomalous couplings in  Eq.~(\ref{eq:nonL}) one by one without and with the kinematic cuts on the $p_{T}^{h,\rm cm}$. In Fig.~\ref{fig:coefs_1} and Fig.~\ref{fig:coefs_2}, we show the linear dependency $A_i$ on the anomalous couplings, as defined in
\begin{equation}
    \sigma (\mu^+ \mu^- \rightarrow h h \nu \bar{\nu}) = \sigma_{\rm SM} \left(1 + A_i C_i + \cdots \right).
\end{equation}
The dashed lines are computed using full  LO matrix element method by MadGraph5,  while the colored bands are computed using the EWA with the PDF scale varied from $Q = \sqrt{s}/2$ to $2\sqrt{s}$.
We find that for $\{C^{h,2h}_0, C^{h,2h}_9, C^{h,2h}_{10}\}$, the two calculations agree very well, but large discrepancy shows  up when turning on the couplings $C^{h,2h}_5$ and  $C^{h,2h}_6$. The reasons behind them are different, as we explain in the following. 

Let's start with the case of $C_5^{h,2h}$:
\begin{equation}
\label{eq:nonLc5}
\aligned
\mathcal{L}_{C_5} =&~  (C^h_5  \frac{h}{v} +  C^{2h}_5 \frac{h^2}{v^2} )\left(W^+_\mu\mathcal{D}^{\mu\nu}W^-_\nu + {\rm h.c.}\right) \endaligned
\end{equation}
The unique feature here is the presence of Lorentz structure $\mathcal{D}^{\mu\nu}W^-_\nu $, which  for the on-shell $W$ is equivalent to the mass term $m_W^2 W^{-\mu}$. This means that EWA approach completely neglects the off-shellness of the gauge bosons. To be more concrete, we can rewrite the $C_5$ interactions by using equation of motion: 
\beq
D^\nu W_{\mu\nu}^a = ig H^\dagger \frac{\sigma^a}{2} \overleftrightarrow{D}_\mu H + g \sum_f \bar{f}_L \frac{\sigma^a}{2} \gamma_\mu f_L
\eeq
The result reads:
\begin{equation}
\aligned
\mathcal{L}_{C_5} =&~  (C^h_5  \frac{h}{v} +  C^{2h}_5 \frac{h^2}{v^2} )\left(2m_W^2 (1+\frac h v)^2 W_\mu^{+}W^{-\mu} + \frac{g}{\sqrt{2}} W^{\mu +} \sum_f \bar{f}_{uL}\gamma_\mu f_{dL} +h.c.+ \mO( W^3))
\right)
 \endaligned
 \label{eq:c5h_AB}
\end{equation}
This is in agreement with the SMEFT operator relation~\cite{Elias-Miro:2013mua,Contino:2013kra}
\beq
\mO_W = g^2\left( -\frac32 \mO_H + 2 \mO_6 + \frac12 (\mO_{y_u} + \mO_{y_d} + \mO_{y_e}) + \frac14\sum_{f = q, l}\mO_L^{(3)f} \right)\\
\eeq
where the contact interactions  $h(h^2)Wf\bar f$ come from the operators $\mO_L^{(3)f}$
\beq
\mO_L^{(3)q} = i\left( H^\dagger\sigma^a  \overleftrightarrow{D}_\mu H \right ) \bar{Q}_L\sigma^a \gamma^\mu Q_L, \quad \mO_L^{(3)l} = i \left( H^\dagger\sigma^a  \overleftrightarrow{D}_\mu H \right ) \bar{L}_L\sigma^a \gamma^\mu L_L.
\eeq

\begin{figure}[tb]
\centering
\begin{subfigure}{0.25\textwidth}
\includegraphics[width=\textwidth]{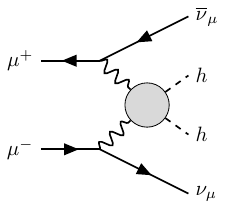}
\caption{}
\end{subfigure}
\hspace{0.5cm}
\begin{subfigure}{0.25\textwidth}
\includegraphics[width=\textwidth]{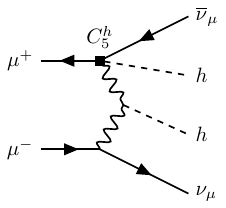}
\caption{}
\end{subfigure}
\hspace{0.5cm}
\begin{subfigure}{0.25\textwidth}
\includegraphics[width=\textwidth]{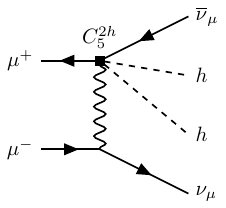}
\caption{}
\end{subfigure}
\caption{\em Representative Feynman diagrams induced by $C_5^h$ and $C_5^{2h}$ for the scattering $\mu^+\mu^-\rightarrow h h \nu_\mu \overline{\nu}_\mu$. }
\label{fig:feyn_c5h}
\end{figure}
Using the equation of motion, the Feynman diagrams that are relevant for $C_5^h$ can be re-arranged accordingly, as shown in Fig.~\ref{fig:feyn_c5h}. Figure~\ref{fig:feyn_c5h}(a) is induced by the first term in Eq.~(\ref{eq:c5h_AB}), which can be captured by the EWA. However, Figs.~\ref{fig:feyn_c5h}(b) and (c), induced by the Higgs-fermion-gauge bosons contact interaction in Eq.~(\ref{eq:c5h_AB}), have very different pole structures in the scattering amplitudes, and therefore, cannot be captured by VBF and EWA at all. In Figs.~\ref{fig:coefs_4} and \ref{fig:coefs_5}, we show the impact of such contact interaction to $C^h_5$ and $C^{2h}_5$ when using the EWA. Once the contribution from the contact interactions is added to the EWA results, a much better agreement is achieved for energy regimes away from the threshold. One also notices that the first term has the same structure as $C_0$ and to be more explicit, by neglecting the second contact term, the coupling $C_5^h$ corresponds to the following identification:
\beq
C_0^h = C_5^h, \qquad C_0^{2h} = 4 C_5^h
\eeq
and the coupling $C_5^{2h}$ corresponds to:
\beq
C_0^{2h} = 2C_5^{2h}
\eeq
This further explains the behaviors of the $C_0,C_5$ combinations in Table~\ref{tab:mhigheL}. 

\begin{figure}[tb]
\centering
\includegraphics[width=.48\textwidth]{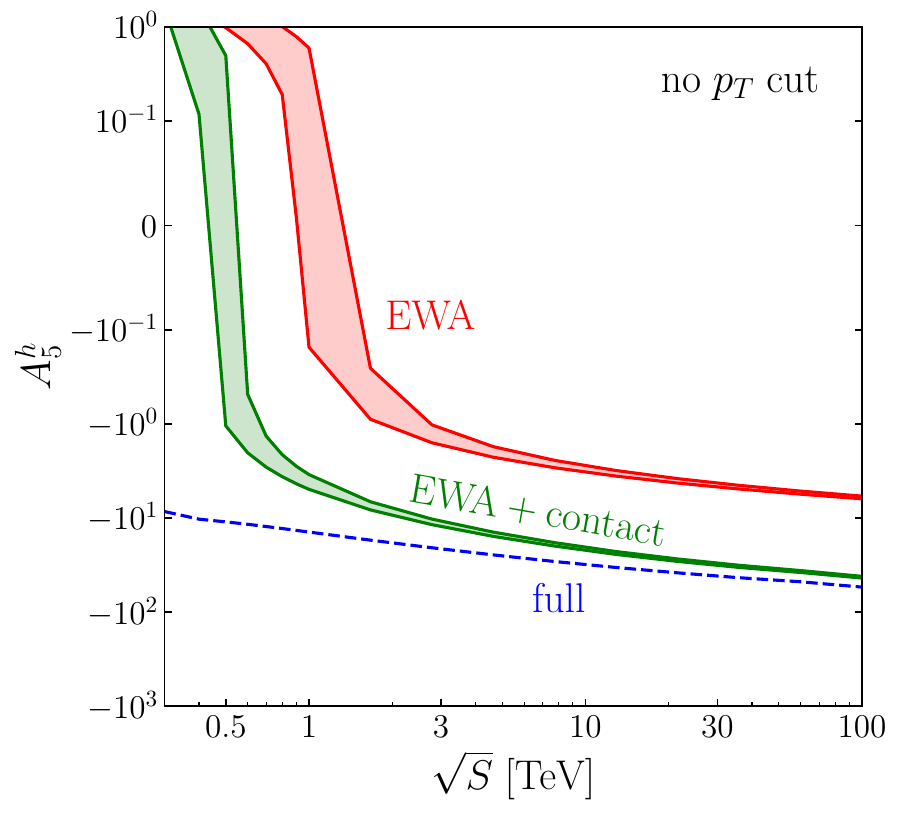}
\includegraphics[width=.48\textwidth]{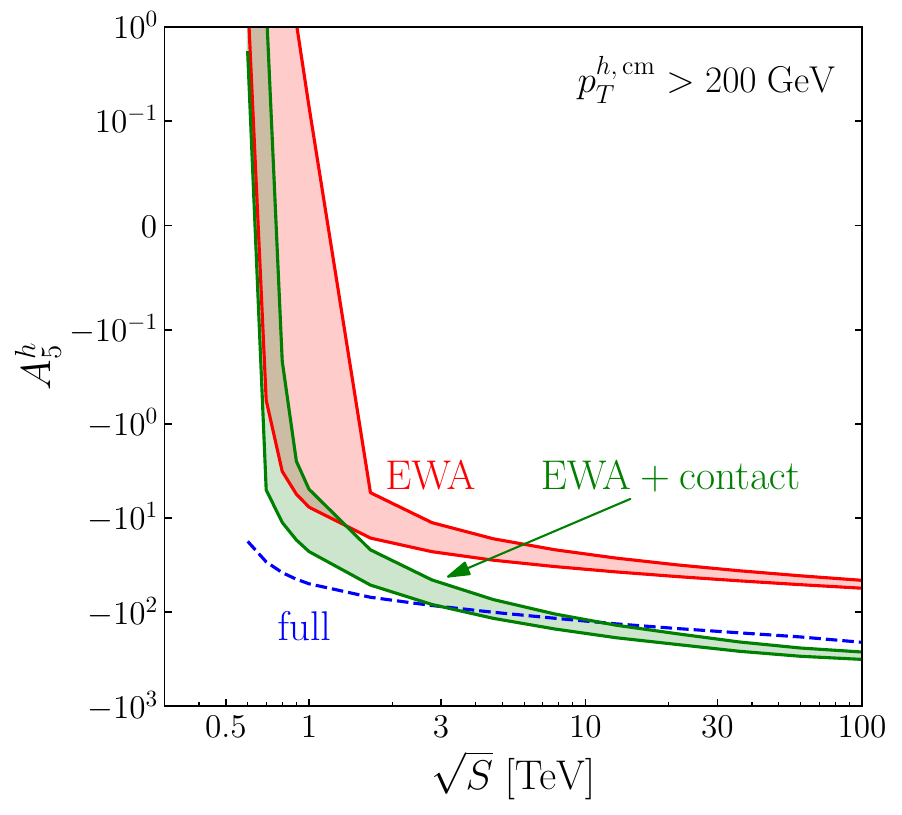}
\caption{\em The linear dependence $A_5^h$ on the anomalous couplings $C_5^h$. The solid red lines are computed using EWA, the solid green lines are the sum of the EWA calculation and the contribution from the Higgs-fermion-gauge boson contact interactions, and the dashed lines come from the full matrix element calculation by MadGraph5. The bands correspond to varying the W PDF scale from $Q = \sqrt{\hat{s}}/2$ to $Q=2\sqrt{\hat{s}}$. Left: No $p_T$ cut applied on the Higgs pair. Right: same as the left plots but with cut on the transverse momentum of the Higgs bosons  $p_T^{h, \rm cm} > 200$ GeV.}
\label{fig:coefs_4}
\end{figure}

\begin{figure}[tb]
\centering
\includegraphics[width=.48\textwidth]{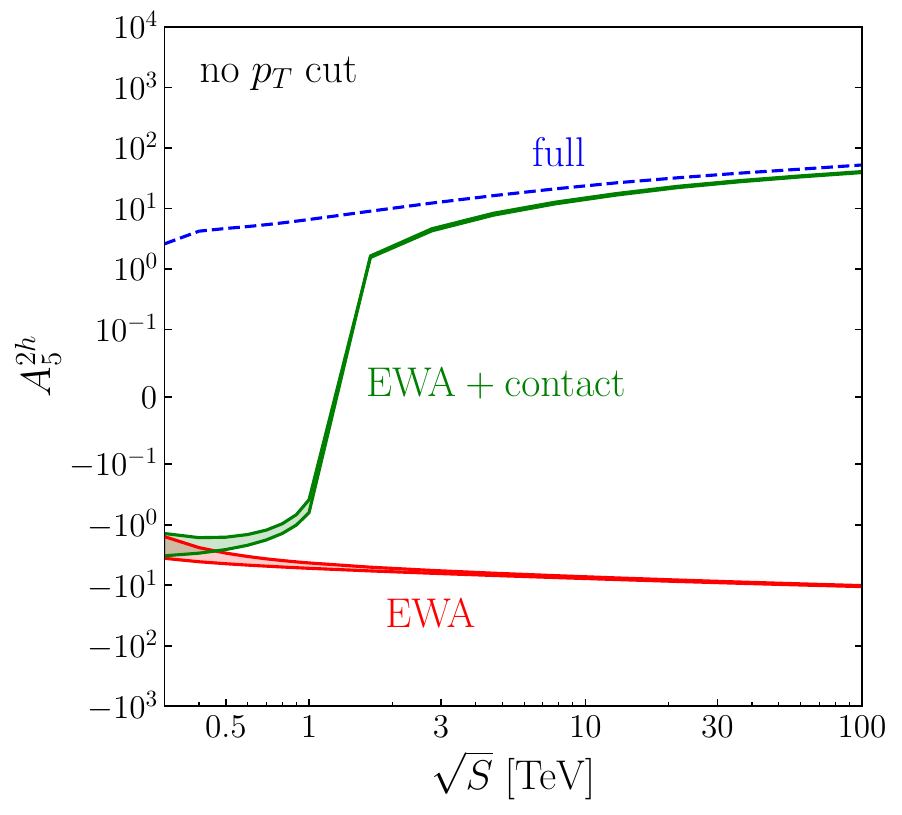}
\includegraphics[width=.48\textwidth]{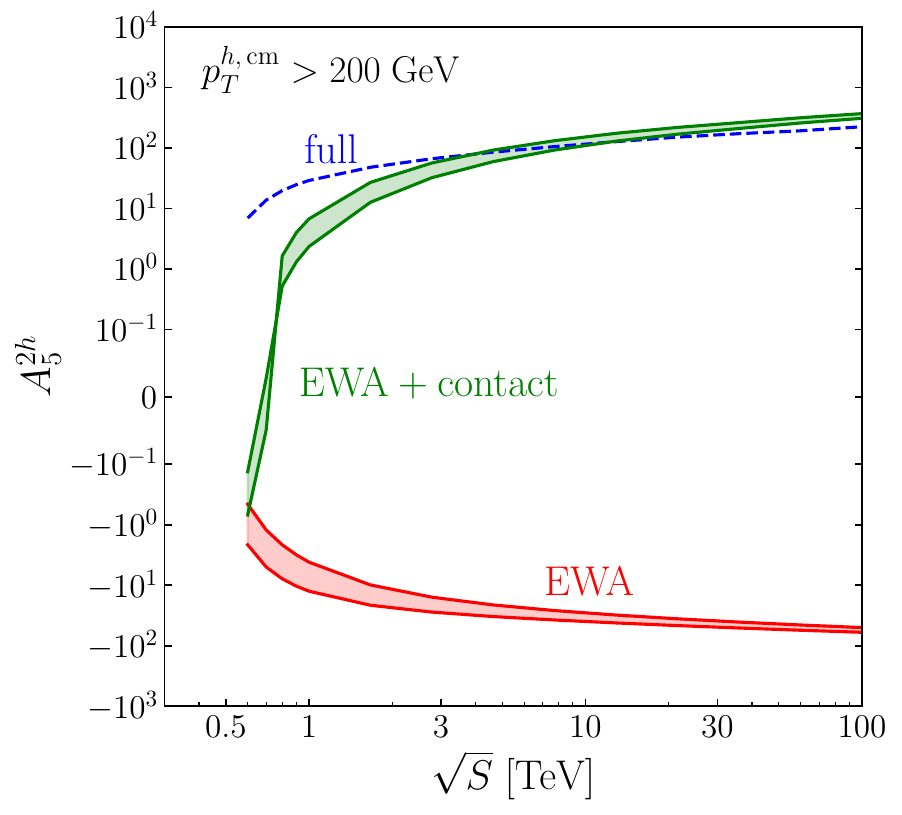}
\caption{\em The linear dependence $A_5^{2h}$ on the anomalous couplings $C_5^{2h}$. The solid red lines are computed using EWA, the solid green lines are the sum of the EWA calculation and the contribution from the Higgs-fermion-gauge boson contact interactions, and the dashed lines come from the full matrix element calculation by MadGraph5. The bands correspond to varying the W PDF scale from $Q = \sqrt{\hat{s}}/2$ to $Q=2\sqrt{\hat{s}}$. Left: No $p_T$ cut applied on the Higgs pair. Right: same as the left plots but with cut on the transverse momentum of the Higgs bosons  $p_T^{h, \rm cm} > 200$ GeV.}
\label{fig:coefs_5}
\end{figure}

Next we consider the case of $C_6$. We expect that the discrepancy comes from terms neglected in EWA. To see this, it is worth recalling the generalized EWA (gEWA) formula introduced in Ref.~\cite{Borel:2012by}, which states that the full amplitude of the $\mu^+ \mu^- \rightarrow h h \nu_\mu \bar \nu_\mu$ can be approximately written as the sum of three on-shell sub-amplitudes:
\beq
\mA_{\rm gEWA} = \frac{4 C_1 C_2}{V_1^2 V_2^2} \sum_{h_1 h_2} \tilde{p}_\bot^{-h_1}  \tilde{q}_\bot^{h_2} g_{-h_1}(x_1) g_{h_2}(x_2)\mA_{\rm on-shell} (W^+_{h_1}(\vec k_1) W^-_{h_2} (\vec k_2) \rightarrow h h) \ ,
\eeq
where $\vec k_{1,2}$ are the three momenta of the $W^\pm$:
\beq
\vec k_1 = (- \vec p_\bot, x_1 E), \qquad \vec k_2 = (- \vec q_\bot, - x_2 E) \ ,
\eeq
where $\vec p_\bot$ and  $\vec q_\bot$ are the transverse momenta of the outgoing neutrinos with respect to the incoming beam. The splitting $g$-functions are given in Eq.~(46) of Ref.~\cite{Borel:2012by}, which are not relevant in the following discussion. We have also abbreviated the helicity-dependent transverse momentum dependence factors as:
\beq
\tilde{p}_\bot^{+1} = \tilde{p}_\bot= p_\bot e^{-i\phi_1} ,\qquad \tilde{p}_\bot^{-1} = \tilde{p}_\bot^*=  p_\bot e^{+i\phi_1}, \qquad \tilde{p}_\bot^{0} = m_W 
\eeq
and similarly for $\tilde{q}_\bot^{h_2} $. Here $\phi_1,\phi_2$ are the azimuthal angles of the outgoing $\bar \nu_\mu, \nu_\mu$ respectively. The inversion of the helicities in the splitting functions for the $W^+$ is due to CP-invariance. Note that  EWA  in Eq.~(\ref{eq:Wpdf}) is obtained by setting $\vec p_\bot, \vec q_\bot$ to zero in the three-momenta of the $W$-bosons and integrating over the azimuthal angles of the outgoing  neutrinos.

 Corrections to EWA can arise when expanding the on-shell amplitudes $\mA_{h_1 h_2}$ in terms of the transverse momenta:
\beq
\mA_{h_1 h_2} (\vec k_1, \vec k_2) \sim \mA_{h_1 h_2}^{(00,00)} + \mA_{h_1 h_2}^{(10,00)} \frac{\tilde p_\bot}{E} + \mA_{h_1 h_2}^{(01,00)} \frac{\tilde p^{*}_\bot}{E}+ + \mA_{h_1 h_2}^{(00,10)} \frac{\tilde q_\bot}{E} + \mA_{h_1 h_2}^{(00,01)} \frac{\tilde q^{*}_\bot}{E} + \cdots
\eeq
After integrating out the azimuthal angles of the outgoing neutrinos, sub-leading terms in the above equation can lead to  interference between the longitudinal polarization and the transverse ones:
\beq
\label{eq:c6gEWA}
\begin{split}
\int_\phi |\mA_{\rm gEWA}|^2 &\sim p_\bot^2 q_\bot^2 |\mA^{(00,00)}_{TT}|^2 + m_W^4 |\mA^{(00,00)}_{LL}|^2 + m_W^2 p_\bot^2 |\mA^{(00,00)}_{TL}|^2 \\
&+ m_W^2 p_\bot^2 \frac{m_W}{E} \mA^{(01,00)}_{LL} \mA^{(00,00)}_{TL} + \cdots
\end{split}
\eeq
where $p_\bot, q_\bot$ are the transverse momenta of the outgoing neutrinos. The last term arises from the interference between the leading term of $\mA_{TL}$ and the sub-leading term of   $\mA_{LL}$.  EWA only keeps the first three terms and treated the last term as higher order corrections.  The last term is usually suppressed by factors of $m_W/E$ compared with the leading terms, as the helicity amplitudes in the SM at tree-level for 2-to-2 scattering are at most constant in the high energy limit. However, this is not true anymore in the presence of anomalous couplings, where energy growing behaviors are expected. 
As can be seen from Table~\ref{tab:leadingS}, in the high energy limit and in the central region, which is relevant for large $p_T^h$ cut, the interference  between SM  and the anomalous coupling contribution is constant over energy in the same helicity components $(0,0),(\pm,\pm),(\pm,0)$  for $C_6^h$ and $(0,0),(\pm,\pm)$  for $C_6^{2h}$. In contrast, the interference between SM $(0,0)$ helicity amplitude and $C_6^h \, (\pm, 0)$  is growing with energy as $\sqrt{s}$, while for SM $(0,0)$ helicity amplitude and $C_6^{2h} \, (\pm, \pm)$ are growing with energy as $s$.  But for $C_{6}^h\  (C_6^{2h})$ , the final (second last) term is in the same order as  the leading terms. Similar observation has been made for the anomalous triple gauge couplings~\cite{Hwang:2023wad}. For other anomalous couplings, similar issues do not appear because the leading energy growing behaviors exist in the $LL$ component. More insights could be obtained if one could  quantify the contribution of these sub-leading corrections to the discrepancy between EWA and full fixed order calculations. We leave this for future studies.

\section{Kinematic Features at a High Energy Muon Collider}
\label{sec:kinematics}

\begin{figure}[t]
\centering
\includegraphics[width=0.48\textwidth]{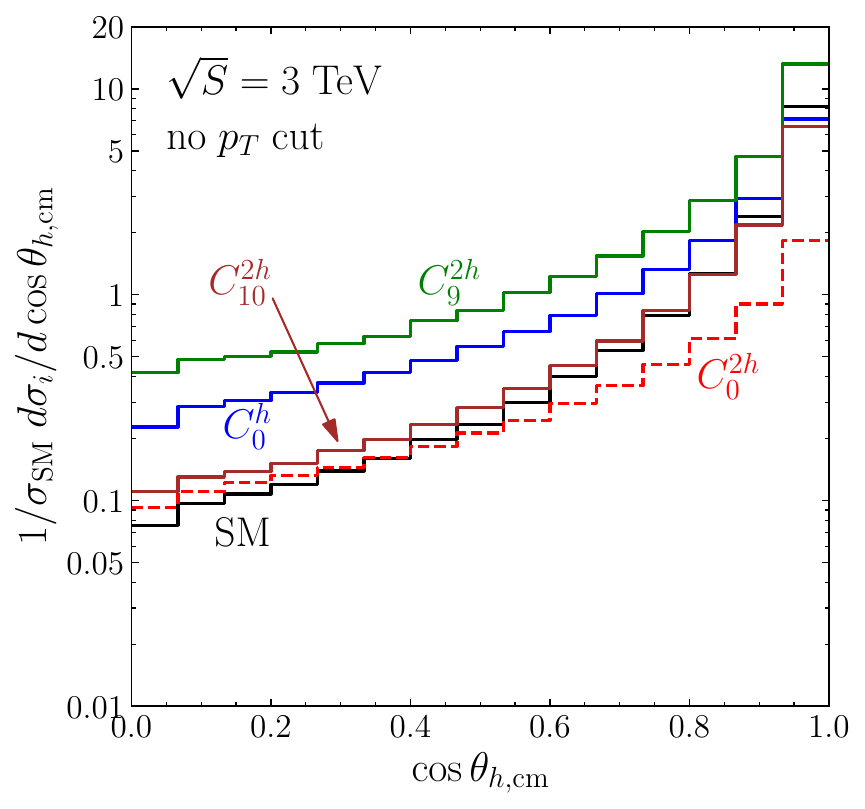}
\includegraphics[width=0.48\textwidth]{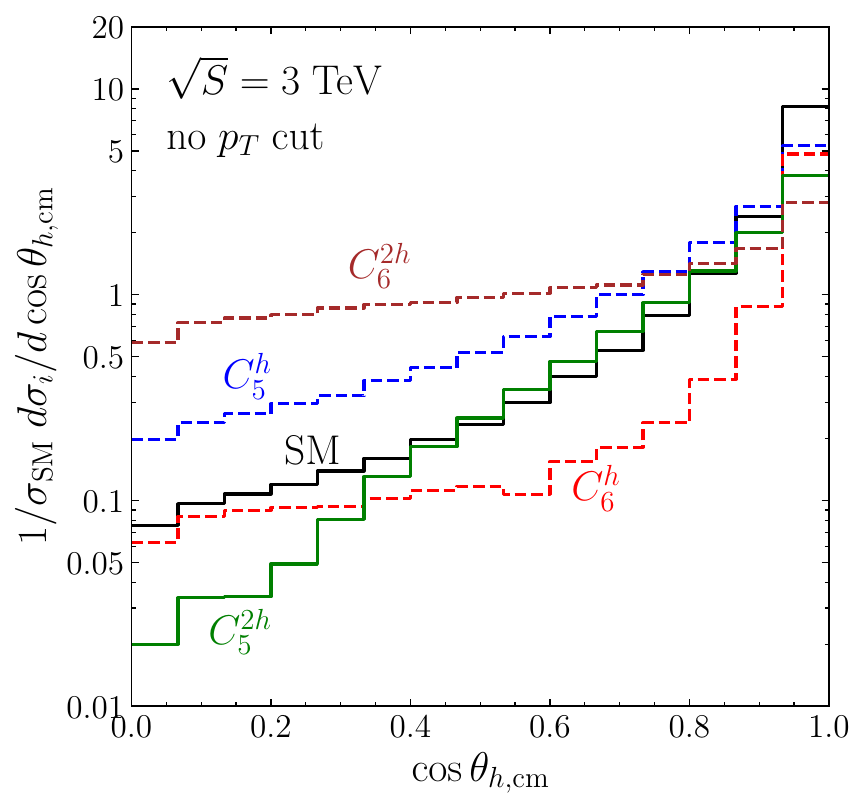}\\
\includegraphics[width=0.48\textwidth]{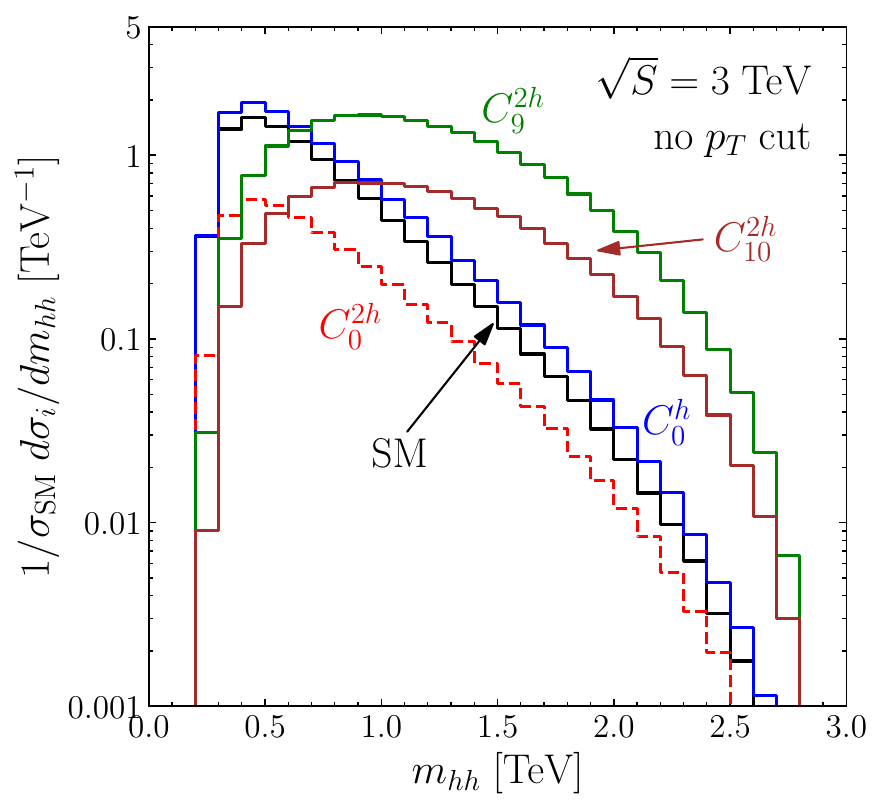}
\includegraphics[width=0.48\textwidth]{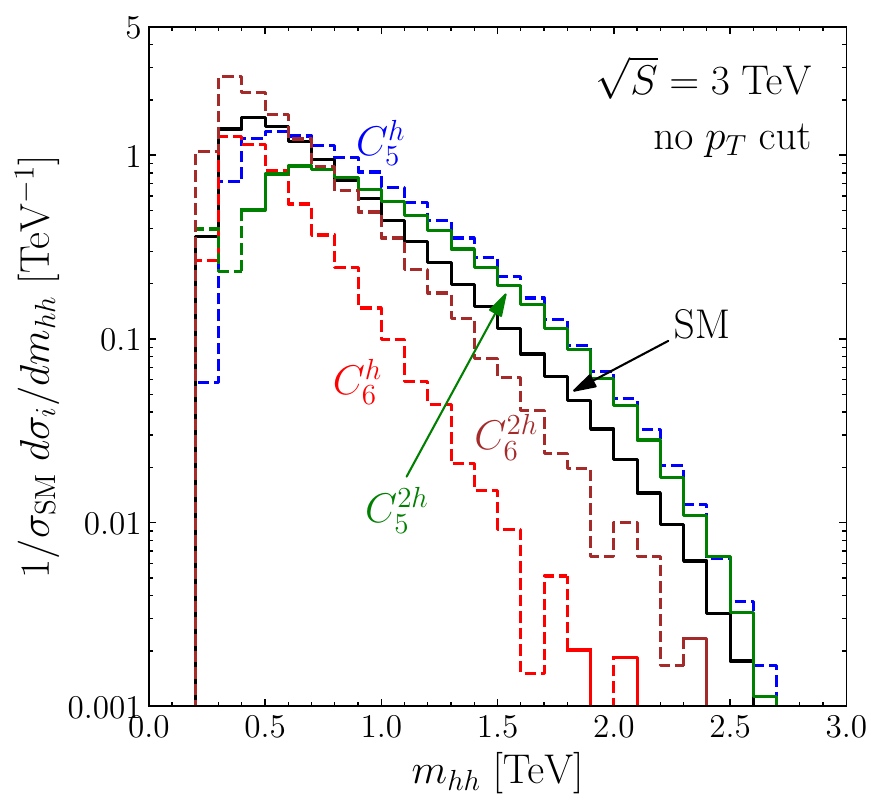}
\caption{\em {$\mu^+ \mu^- \rightarrow h h \nu_\mu \bar \nu_\mu$:} 
$\cos\theta_{h,\rm cm}$ distributions (top panels) and $m_{hh}$ distributions (bottom panels) for the SM and BSM { linear interference contribution} at the 3 TeV muon collider. { The SM distributions are shown in black lines. }All the lines are normalized to the SM total cross section. Dashed lines indicate negative values for destructive interference. ($C_{0}^{h}, C_{0}^{2h}, C_{5}^{h}, C_{6}^{h}, C_{5}^{2h}, C_{6}^{2h}, C_{9}^{2h}, C_{10}^{2h}) = (0.1, 0.1, 0.05, 0.5, 0.05, 0.3, 0.01, 0.01)$. }
\label{fig:kd3TeV}
\end{figure}

\begin{figure}[t]
\centering
\includegraphics[width=0.48\textwidth]{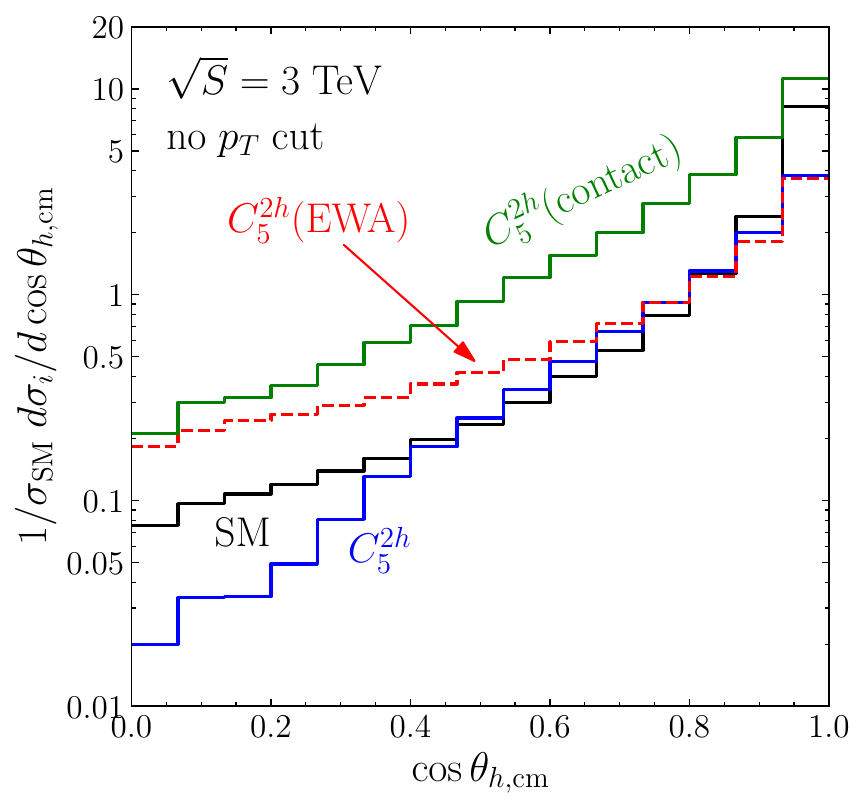}
\caption{\em {$\mu^+ \mu^- \rightarrow h h \nu_\mu \bar \nu_\mu$:} demonstration of the $\cos\theta_{h,\rm cm}$ distribution for $C_5^{2h}$.}
\label{fig:special}
\end{figure}

\begin{figure}[t]
\centering
\includegraphics[width=0.48\textwidth]{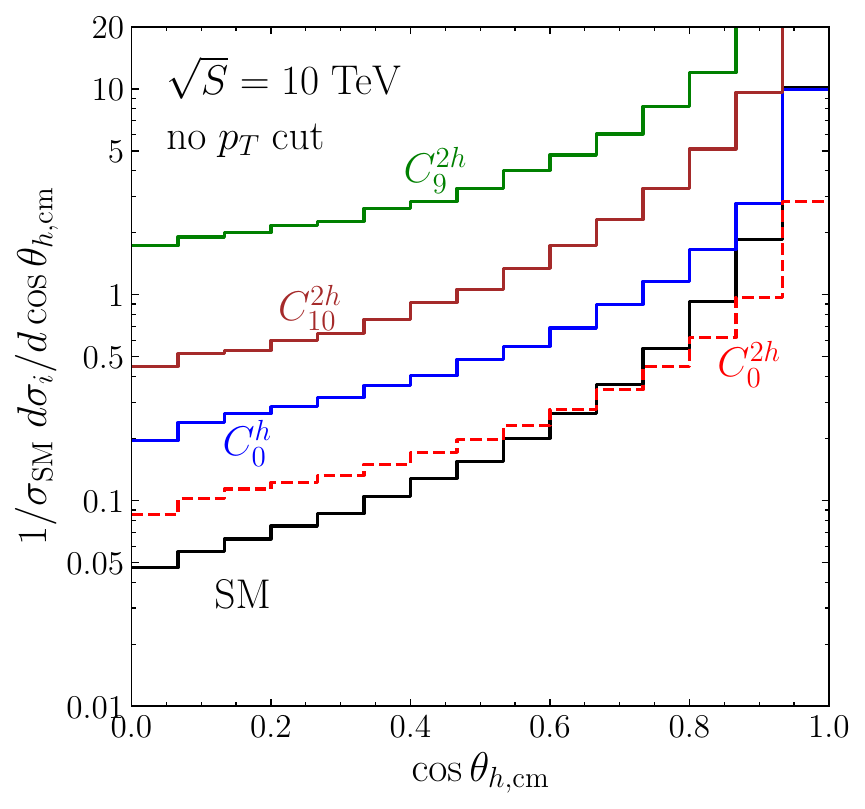}
\includegraphics[width=0.48\textwidth]{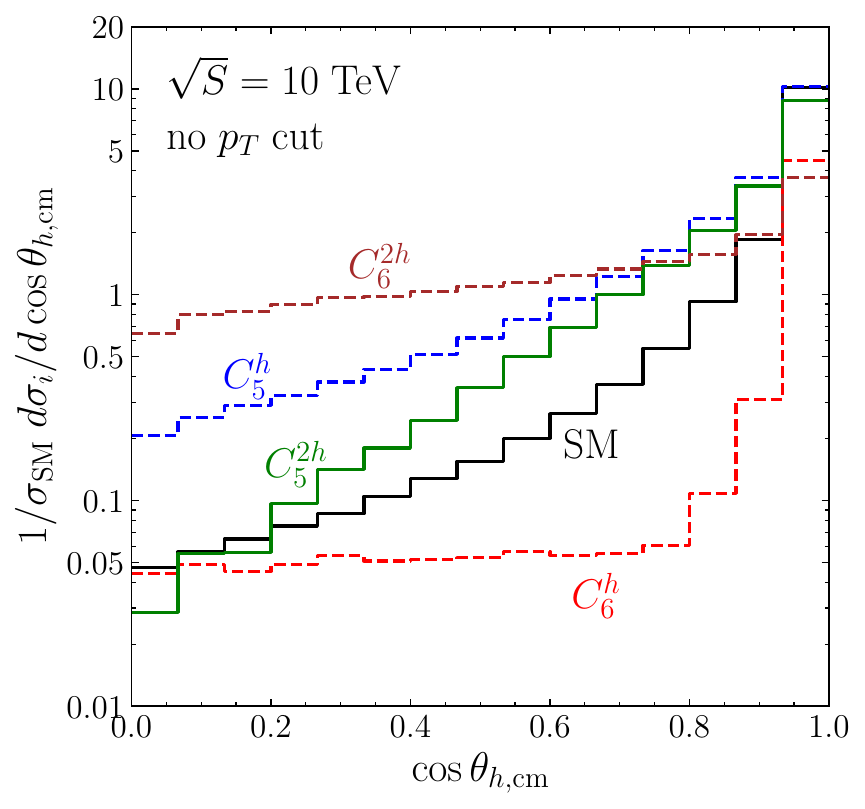}\\
\includegraphics[width=0.48\textwidth]{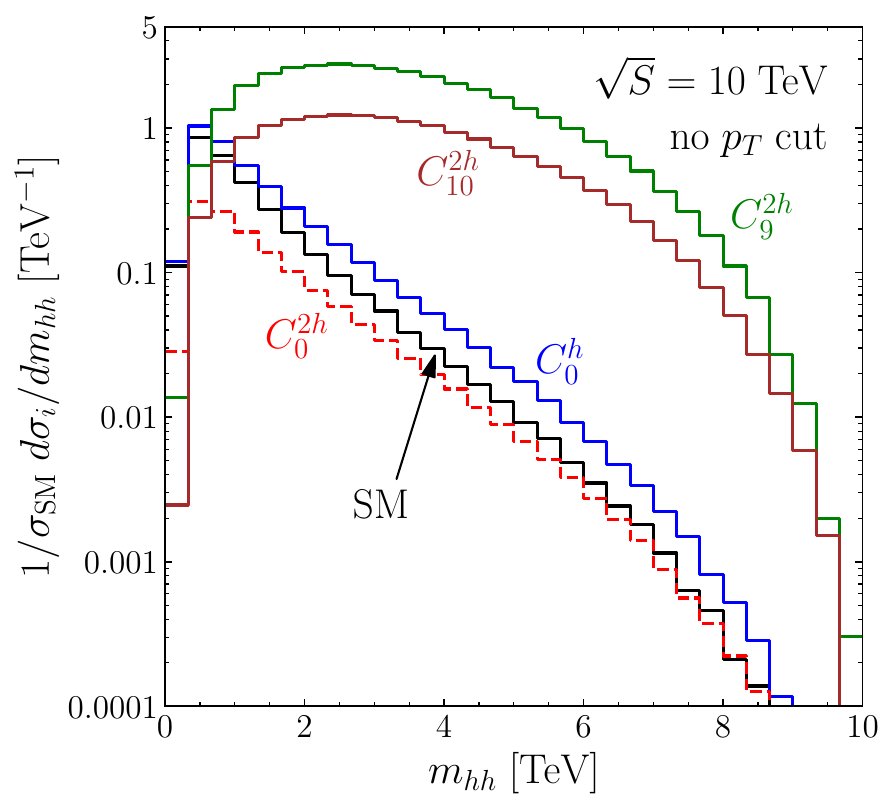}
\includegraphics[width=0.48\textwidth]{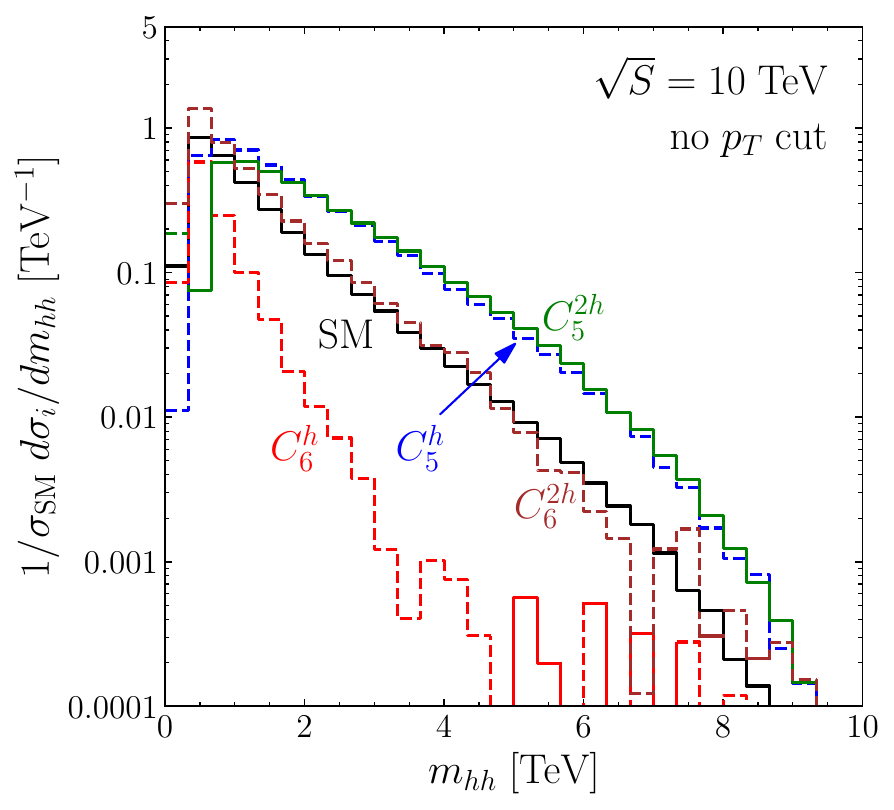}
\caption{\em {$\mu^+ \mu^- \rightarrow h h \nu_\mu \bar \nu_\mu$:} 
$\cos\theta_{h,\rm cm}$ distributions (top panels) and $m_{hh}$ distributions (bottom panels) for the SM and BSM linear interference contribution at a 10 TeV muon collider. The SM distributions are shown in black lines. All the lines are normalized to the SM total cross section. Dashed lines indicate negative values for destructive interference. ($C_{0}^{h}, C_{0}^{2h}, C_{5}^{h}, C_{6}^{h}, C_{5}^{2h}, C_{6}^{2h}, C_{9}^{2h}, C_{10}^{2h}) = (0.1, 0.1, 0.05, 0.5, 0.05, 0.3, 0.01, 0.01)$. }
\label{fig:kd10TeV}
\end{figure}
 
In this section, we  present kinematic distributions for double Higgs production  at the 3 TeV and 10 TeV muon colliders, both with and without the presence of anomalous couplings. A detailed collider study on the expected sensitivity on constraining the anomalous couplings is beyond the scope of the present work. 

In Fig.~\ref{fig:kd3TeV}, we  plot the distributions of $\cos \theta_{h, \rm cm }$, where $\theta_{h, \rm cm }$ is  the  scattering angle in the partonic CM frame of the two Higgs bosons, and $m_{hh}$, the invariant mass of the two Higgs bosons, at  the 3 TeV muon collider without any kinematic cuts, normalized to the SM cross section. The simulation is performed using MadGraph5 at the leading order (LO). To highlight the difference between the case of SM and that of anomalous couplings, we are only showing leading contributions from the interference terms $d\sigma_i$ which linearly depends on the anomalous couplings,
\beq
d\sigma_{\rm BSM} \equiv d\sigma_{SM} + d\sigma_i(C_i) + \mathcal{O}(C_i^2,\, C_i C_j).
\eeq 
Dashed lines indicate negative values for destructive interference. The following values for the anomalous couplings are chosen as benchmark points:
\beq
(C_{0}^{h}, C_{0}^{2h}, C_{5}^{h}, C_{6}^{h}, C_{5}^{2h}, C_{6}^{2h}, C_{9}^{2h}, C_{10}^{2h}) = (0.1, 0.1, 0.05, 0.5, 0.05, 0.3, 0.01, 0.01)\ .
\eeq
We show the results for $C_{0,9,10}$ on the left panels, and $C_{5,6}$ on the right panels.
We see clearly from the distributions of $\cos \theta_{h,\rm cm}$ that the presence of the anomalous couplings tend to make two Higgs bosons more central. As can be seen from Table~\ref{tab:mhigheL}, this is due to the fact that the energy growing pieces of the interference terms only have at most one power of $t/u$-channel singularity ((0,0) helicity configuration), while for the SM cross section,  $t/u$-channel singularity is at the second power. The only exception is $C_5^{2h}$, where  the distribution is suppressed in the central region compare with SM. As demonstrated in Fig.~\ref{fig:special}, this is due to the cancellation between two contributions: EWA, which is obtained by convolving the $W$-PDF with the helicity amplitudes calculated in Section~
\ref{sec:helamp}, and the contact term contribution, which is obtained by implementing the $\mu \nu_\mu W h h $ vertex in Eq.~(\ref{eq:c5h_AB}).

Determining the kinematic distributions for the $hh$ production is of fundamental importance in exploring the underlying dynamics within or beyond the SM. 
From the $m_{hh}$ distributions  in Fig.~\ref{fig:kd3TeV}, we examine the energy growing behavior of various anomalous couplings at the linear order of the interference with the SM contributions, 
\beq \label{eq:c56sca}
\dfrac{d\sigma_{i}/dm_{hh}}{d\sigma_{\rm SM}/dm_{hh}}
\sim \left\{\begin{array}{cccc}
m_{hh}^2 \ln m_{hh}& \quad C_{9}^{2h}, C_{10}^{2h}     \\
\ln m_{hh}   & \quad C_{0}^{h,2h}  \\
\left(\ln m_{hh}\right)^2   & \quad C_5^{h,2h}  \\
 1/m_{hh}^2\, \ln m_{hh}  & \quad C_{6}^{h,2h} \ ,
\end{array}
\right.
\eeq
As discussed in the previous section, the scaling behaviors for $C_0^{h,2h}$ and $C_{9,10}^{2h}$ can be understood within the EWA, where the di-Higgs invariant mass $m_{hh}$ is given by the partonic center-of-mass energy $\sqrt{s}$:
\beq
\dfrac{d\sigma_{i}/dm_{hh}}{d\sigma_{\rm SM}/dm_{hh}}
\sim \dfrac{\hat \sigma_{0,0\ \rm BSM}(\sqrt{s}=m_{hh})}{\hat \sigma_{0,0\ \rm SM}(\sqrt{s}=m_{hh})},
\eeq
and the partonic cross section behaviors can be obtained from Eq.~(\ref{eq:partxshe}). Note that the energy $\sqrt{s}$ dependence in the EW PDFs cancels because both the SM contribution and the BSM contribution induced by $C_0^{h,2h}$ and $C_{9,10}^{2h}$ are dominated by the scattering of the longitudinally polarized $W$ bosons. The enhancement at higher $\sqrt{s}=m_{hh}$ is due to the nature of higher-dimensional operators, and the logarithmic factor comes from the $t$- and $u$-channel singularity in the interference terms, as discussed in Sec.~\ref{sec:HEL}.
Although the EWA provides an intuitive understanding of the scaling behaviors for $C_0^{h,2h}$ and $C_{9,10}^{2h}$, it does not necessarily capture the correct  scaling behaviors for $C_5^{h,2h}$ and $C_6^{h,2h}$,  which have distinct energy dependence, as pointed out in Sec.~\ref{sec:ewabsm} and given in Eq.~(\ref{eq:c56sca}). Therefore, measuring the energy spectrum of the $m_{hh}$ distribution would shed light on the underlying physics for different operators. Similar behaviors are also observed at higher colliding CM energies. We illustrate those in Fig.~\ref{fig:kd10TeV} for a 10 TeV muon collider, where we have also plotted the distributions of $\cos\theta_{h,\rm cm}$ at 10 TeV.

\section{Discussions and Conclusion}
\label{sec:conclusion}

One of the unique roles of the SM Higgs boson  lies in the fact that  its couplings to other SM particles unitarize the energy growing behaviors in high energy scatterings involving  gauge bosons and the Higgs boson.  Studying these processes could potentially reveal the microscopic nature of the Higgs boson, whether it belongs to an electroweak doublet  and whether it is an elementary scalar or a composite particle \`a la a pseudo-Nambu-Goldstone boson. Among the relevant processes, $V V \rightarrow hh $ offers an additional opportunity to measure $VVhh$ coupling and trilinear Higgs coupling, both of which have not been verified experimentally and have posed significant challenges for future experiments. Moreover, when considering possible modifications to the SM Higgs couplings, it is important to keep open the possibility that some new coupling structures beyond those in the SM may show up and have a sizable presence, either due to accidentally large contributions from higher dimensional operators or the nonlinear structure of the EFT.

In this paper, we studied in detail the very high energy scattering of electroweak particles, which is dominated by vector boson fusion and directly probes the origin of electroweak symmetry breaking. Using the production of the Higgs pair  at a multi-TeV muon collider as the prime example, $\mu^+ \mu^- \rightarrow h h \nu_\mu \bar \nu_\mu$, we critically examine the validity of the EWA/EW PDF approach versus a full fixed order calculation in the presence of anomalous couplings defined in Eq.~(\ref{eq:nonL}). An important feature of our study is to   treat the single Higgs anomalous couplings $VVh$  as  independent from the double-Higgs anomalous couplings $VV hh$, and also include  new Lorentz structures up to two derivatives  on the fields.

To facilitate the comparison between the EWA and the full fixed-order calculations, we computed in detail  the helicity amplitudes of the subprocess $W^+ W^- \rightarrow h h$ in the presence of the  anomalous $WW h$ and $WW h h$ couplings, paying particular attention to  the threshold behaviors and in the high energy limits. 
We found  agreements between the two formulisms in the couplings $C_0^{h,2h}$ and $C_{9,10}^{2h}$, where there is no derivative acting on the electroweak gauge bosons and the Lorentz structures do not deviate {qualitatively} from those in the SM. However, when there are derivatives acting on the gauge bosons, we found significant discrepancies between EWA/EW PDF and the full fixed-order calculations in the anomalous couplings $C_5^{h,2h}$ and $C_6^{h,2h}$. In the former case we identified the reason behind the discrepancies as a missing contact term in Eq.~(\ref{eq:c5h_AB}), which is not captured by EWA. In the latter case, we observe that the difference arises from  ``subleading'' interference terms between different helicity configurations as in Eq.~(\ref{eq:c6gEWA}). Our findings suggest, in order to study effects of anomalous Higgs couplings in very high energy electroweak scatterings, the employment of electroweak PDFs should be dealt {with care to account for the potential discrepancies} discovered in the present study. {It boils down to two essential points: 
\begin{itemize}
    \item The EWA/EW PDF approach relies on the collinear factorization of the high energy scattering in the SM, which may miss important contributions due to certain new 4-point interactions $C_5^{h,2h}$;
    \item The EWA/EW PDF approach performs an incoherent sum of different initial state helicity contributions, which may miss certain sizable interference effects from off-shell non-collinear $W$'s $C_6^{h,2h}$.
\end{itemize}
}
\noindent 
We hope to come up with concrete implementations {to capture the characteristic features of the new physics} in a future work.

We also presented the distributions of  kinematic variables, the cosine of the scattering angle of the Higgs boson in the partonic center-of-mass frame $\cos\theta_{h,cm}$ and the invariant mass of the di-Higgs system $m_{hh}$. We include effects of anomalous Higgs couplings in the LO simulations of the full process $\mu^+ \mu^- \rightarrow h h \nu \bar \nu$. We found that, with the exception of $C_5^{2h}$,  anomalous couplings  tend to make the distribution of the scattering angle more central. In addition, there are energy growing behaviors observed in the inclusive cross section for the anomalous couplings $C_{9,10}^{2h}$. {We reiterate the importance to study the detailed kinematical distributions in  probing the underlying BSM physics.} These will provide insights into future studies on the prospects for measuring these anomalous couplings at a very high energy muon collider and hadron collider.

\begin{acknowledgments}
 The work of DL and TH was supported in part by the U.S.~Department of Energy under grant No.~DE-SC0007914 and in part by the PITT PACC. The work of DL was also supported by DOE Grant Number DE-SC-0009999. This work is supported in part by the U.S. Department of Energy under contracts No.~DE-SC0010143. Work at Argonne is supported in part by the U.S. Department of Energy under contract DE-AC02-06CH11357. The work of XW was supported in part by the National Science Foundation under Grant No.~PHY-2210177.
 \end{acknowledgments}

\appendix

\section{Feynman Rules for the Higgs-$W$ couplings}
\label{appendix:B}
The relevant Feynman rules for the $hhh$, $WWh$, and $WWhh$ vertices are given below, 
\begin{flalign}
        \begin{gathered}\includegraphics[width = 0.2\textwidth]{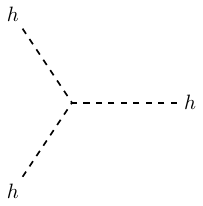}\end{gathered} & \begin{aligned}=~ -6i\lambda v,\end{aligned} \\
        \begin{gathered}\includegraphics[width = 0.2\textwidth]{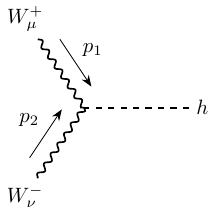}\end{gathered} & \begin{aligned}&\vphantom{\dfrac{2m_W^2}{v}}\\=~ &i\dfrac{2m_W^2}{v}\left(1+C^h_0\right)\eta^{\mu\nu}+i\dfrac{C^h_5}{v}\left[\left(p_1^2+p_2^2\right)\eta^{\mu\nu}-p_1^\mu p_1^\nu-p_2^\mu p_2^\nu\right] \\
        &+i\dfrac{2C^h_6}{v}\left[p_1^\nu p_2^\mu - \left(p_1\cdot p_2\right) \eta^{\mu\nu}\right],\end{aligned} \\
        \begin{gathered}\includegraphics[width = 0.2\textwidth]{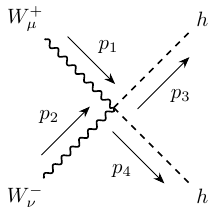}\end{gathered} & \begin{aligned}&\vphantom{\dfrac{2m_W^2}{v}}\\&\vphantom{\dfrac{2m_W^2}{v}}\\
        =~&i\dfrac{2m_W^2}{v^2}\left(1+C^{2h}_0\right)\eta^{\mu\nu}+i\dfrac{2C^{2h}_5}{v^2}\left[\left(p_1^2+p_2^2\right)\eta^{\mu\nu}-p_1^\mu p_1^\nu-p_2^\mu p_2^\nu\right]\\
        &+i\dfrac{4C^h_6}{v^2}\left[p_1^\nu p_2^\mu - \left(p_1\cdot p_2\right) \eta^{\mu\nu}\right]-i\dfrac{2C^{2h}_9}{v^2}\left(p_3\cdot p_4\right)\eta^{\mu\nu}\\
        &-i\dfrac{C^{2h}_{10}}{v^2}\left(p_3^\mu p_4^\nu+p_3^\nu p_4^\mu\right).\end{aligned}
\end{flalign}

\section{Reduced helicity amplitudes for $W^+W^-\to hh$}
\label{appendix:A}

In this appendix, we collect the full formulae for the reduced helicity amplitudes defined in Eq.~(\ref{eq:modifiedMs}) for the subprocess $W^+W^-\to hh$.

\begin{alignat}{2}
\mathcal{A}^s_{0,0} =&~\frac{12\lambda}{g^2}\left[(1+\beta_W^2)(1 + C_0^h + C^h_5) - C^h_6\gamma_W^{-2}\right], \\
\mathcal{A}^t_{0,0} =& ~\left[-2(1+\beta_W^2) - 2\gamma_W^2(\beta_W-\beta_h\cos\theta)^2\right](1+2C_0^h)\nonumber \\
&\quad + 2C^h_5\left[-(1+\beta_W^2)\frac{m_h^2}{m_W^2} +2\beta_W\beta_h\cos\theta+ 2\gamma_W^2(\beta_W^2-\beta_h\cos\theta)(\beta_W^2+\beta_h\cos\theta)\right]\nonumber\\
&\quad -4C^6_h(\beta_W-\beta_h\cos\theta)\ , \\
\mathcal{A}^u_{0,0} =&~ \left[-2(1+\beta_W^2) - 2\gamma_W^2(\beta_W+\beta_h\cos\theta)^2\right](1+2C^h_0)\nonumber \\
&\quad + 2C^h_5\left(-(1+\beta_W^2)\frac{m_h^2}{m_W^2} -2\beta_W\beta_h\cos\theta+ 2\gamma_W^2(\beta_W+\beta_h\cos\theta)(\beta_W^2-\beta_h\cos\theta)\right)\nonumber\\
&\quad -4C^6_h(\beta_W+\beta_h\cos\theta)\ ,\\
\mathcal{A}^4_{0,0} =&~ -4C^{2h}_6 + 2\gamma_W^2(1+\beta_W^2)(1 + C_0^{2h}+ 2C^{2h}_5) \nonumber \\
&\quad - 2\gamma_W^4\left[ C^{2h}_9(1+\beta_W^2)(1+\beta_h^2) + C^{2h}_{10}(\beta_W^2+\beta_h^2\cos^2\theta) \right]\ ,\\
\mathcal{A}^s_{\pm,0} =&~ \mathcal{A}^s_{0,\mp} = 0 \ ,\\ 
\mathcal{A}^t_{\pm,0} =&~ \mathcal{A}^t_{0,\mp} = -2\gamma_W\beta_h(\beta_W-\beta_h\cos\theta)(1 + 2 C_0^h + 2C^h_5) - 4C^h_6\gamma_W\beta_W\beta_h\ ,\\
\mathcal{A}^u_{\pm,0}=&~\mathcal{A}^u_{0,\mp} = 2\gamma_W\beta_h(\beta_W+\beta_h\cos\theta)(1 + 2 C_0^h + 2C^h_5) + 4C^h_6\gamma_W\beta_W\beta_h\ , \\
\mathcal{A}^4_{\pm,0} =&~ \mathcal{A}^4_{0,\mp} =~ 2C^{2h}_{10} \gamma_W^3\beta_h^2 \cos\theta\ ,\\
\mathcal{A}^s_{\pm,\mp}=&~0\ , \\
\mathcal{A}^t_{\pm,\mp} =&~  \mathcal{A}^u_{\pm,\mp}=-\frac{4}{\sqrt{6}}\beta_h^2(1 + 2 C_0^h + 2C^h_5)\  ,\\ 
\mathcal{A}^4_{\pm,\mp} =&~ -\frac{4}{\sqrt{6}} C^{2h}_{10} \gamma_W^2\beta_h^2 \ ,\\
\mathcal{A}^s_{\pm,\pm} =&~-\frac{12\lambda}{g^2}\left[\gamma_W^{-2}(1+C_0^h+C^h_5) - C^h_6(1+\beta_W^2)\right]\ ,\\
\mathcal{A}^t_{\pm,\pm} =&~  (2\gamma_W^{-2}+\beta_h^2\sin^2\theta)(1+2C_0^h) - 2C^h_5\left[-\gamma_W^{-2} + (\beta_W-\beta_h\cos\theta)^2\right] \nonumber\\
&\quad - 4C^h_6(\beta_W-\beta_h\cos\theta)\ , \\
\mathcal{A}^u_{\pm,\pm} =&~  (2\gamma_W^{-2}+\beta_h^2\sin^2\theta)(1+2C_0^h) - 2C^h_5\left[-\gamma_W^{-2} + (\beta_W+\beta_h\cos\theta)^2\right] \nonumber\\
&\quad - 4C^h_6(\beta_W+\beta_h\cos\theta)\ , \\
\mathcal{A}^4_{\pm,\pm} =&~ -2(1+C_0^{2h}) - 4 C^{2h}_5 \nonumber\\
 &\quad + \gamma_W^2\left[ 4C^{2h}_6(1+\beta_W^2) + 2C^{2h}_9(1+\beta_h^2) + C^{2h}_{10}\beta_h^2\sin^2\theta \right]\ .
\end{alignat}


\bibliography{NonLinearHiggsRefs}
\bibliographystyle{utphys}

\end{document}